\documentclass[pre,aps,twocolumn,superscriptaddress,notitlepage,nofootinbib]{revtex4-1}
\usepackage{amssymb,epsfig,amsmath,bm,subfigure,graphicx,textcomp,url,float,color,cases}

\usepackage{epsfig}
\usepackage{textcomp}
\usepackage[normalem]{ulem}

\usepackage[colorlinks=true, urlcolor=blue, anchorcolor=blue, citecolor=blue,filecolor=blue,linkcolor=blue,menucolor=blue]{hyperref}

\usepackage{color}

\usepackage[titletoc,title]{appendix}

\newcommand{\ii}{{\rm i}}

\begin{document}

\title{Finite-time blowup of a Brownian particle in a repulsive potential}

\author{P. L. Krapivsky}
\affiliation{Department of Physics, Boston University, Boston, Massachusetts 02215, USA}
\affiliation{Santa Fe Institute, Santa Fe, New Mexico 87501, USA}
\author{Baruch Meerson}
\affiliation{Racah Institute of Physics, Hebrew University of Jerusalem, Jerusalem 91904, Israel}

\begin{abstract}
We consider a Brownian particle performing an overdamped motion in a power-law repulsive potential. If the potential grows with the distance faster than quadratically, the particle escapes to infinity in a finite time. We determine the average blowup time and study the probability distribution of the blowup time. In particular, we show that the long-time tail of this probability distribution decays purely exponentially, while the short-time tail exhibits an essential singularity. These qualitative features turn out to be quite universal, as they occur for all rapidly growing power-law potentials in arbitrary spatial dimensions.  The quartic potential is especially tractable, and we analyze it in more detail.
\end{abstract}

\maketitle

\section{Introduction}
\label{intro}

Finite-time singularities, or blowups, in nonlinear systems are ubiquitous, and they have a special place in physics and mathematics, see, e.g.,  \cite{Mather,Caflisch,Eggers09}. Remarkably, a finite-time blowup occurs already in a simple nonlinear differential equation
\begin{equation}
\label{DE-n}
\dot{x}(t) = g x^n(t)\,,\quad g>0\,,
\end{equation}
where $n>1$ is an integer. The solution of this equation,
\begin{equation}
\label{determ}
x(t) = \frac{x_0}{\left(1-g \nu x_0^{\nu} t\right)^{1/\nu}},\quad \nu\equiv n-1,
\end{equation}
blows up in a finite time $(g \nu x_0^{\nu})^{-1}$ for any positive initial condition: $x(0)\equiv x_0 > 0$.

Equation~(\ref{DE-n}) describes an overdamped deterministic motion of a particle in a repulsive potential
\begin{equation}\label{potential1d}
U(x,n) = -\frac{g x^{n+1}}{n+1}
\end{equation}
in one dimension, and this is probably the simplest possible model of a finite-time blowup. A natural extension of this simple model accounts for noise. In the presence of additive white Gaussian noise, Eq.~(\ref{DE-n}) gives way to the Langevin equation
\begin{equation}
\label{SDE-n}
\dot{x}(t) = gx^n(t)+ \eta(t)\,,\quad g>0\,, n=2,3, \ldots \,,
\end{equation}
with $\langle \eta(t)\rangle =0$ and $\langle \eta(t_1)\eta(t_2)\rangle = 2 D\delta(t_1-t_2)$. Equation~(\ref{SDE-n}) describes an overdamped motion of a Brownian particle in the repulsive potential. Here the blowup time -- the first passage time to infinity -- is a random quantity, and it is interesting to study its statistical properties. With the noise present, the particle escapes to infinity with probability $1$ even for $x_0\leq 0$. For odd $n>1$, the finite-time escapes to $+\infty$ and $-\infty$ are feasible and, for $x_0=0$, equally probable. For even $n\geq 2$, the escape is only to $+\infty$.

The $d$-dimensional Langevin equation
\begin{equation}
\label{SDE-nd}
\dot{\mathbf{r}} = gr^{n-1} \mathbf{r}+ \boldsymbol{\eta}\,,\quad g>0\,,
\end{equation}
describes a Brownian particle in an isotropic repulsive power-law potential
$U(r)= - g r^{n+1}/(n+1)$, where $r=|\mathbf{r}|$.
Here $\boldsymbol{\eta}$ is Gaussian white noise in $d$ dimensions.  A finite-time blowup can occur here for any real $n>1$ and for any $\mathbf{r}(0)\equiv\mathbf{r_0}$.

Intricate interplay of finite-time singularities and additive noise  in dynamical systems  already received attention in the past. References \cite{Bray2000,Farago2000,Fogedby2002} studied such a problem in the context of the Langevin equation
\begin{equation}
\label{LangevinF}
 \dot{x}(t) = -|x(t)|^{-1-\beta}+\eta(t)\,,
\end{equation}
where $\beta\geq 0$ and $\eta(t)$ is a Gaussian white noise. In the absence of the noise, $x(t)$ vanishes at a finite time $t_0$, with a power-law behavior determined by $\beta$, and $\dot{x}$ diverges at $t=t_0$, so the solution does not exist beyond $t=t_0$. In the presence of noise, the time of singularity becomes a random quantity, and one is interested in its statistics. In spite of the obvious similarity  between the problems defined by Eqs.~(\ref{SDE-n}) and (\ref{LangevinF}),  there is an important difference which stems from the fact that, in the case of Eq.~(\ref{SDE-n}), the ``target" is at infinity.

Much closer to our work are Refs.~\cite{Ornigotti,SilerPRL,Ryabov} where Eq.~(\ref{SDE-n}) was studied for $n=2$ and $g<0$. In that case the particle escapes to $-\infty$ rather than to $+\infty$, but otherwise the two models are equivalent. Theoretical predictions made in Ref.~\cite{Ornigotti} (see also \cite{Ryabov}) were verified in experiment with Brownian particles moving near an inflection point in an unstable cubic optical potential \cite{SilerPRL}. 
(The earlier stage of the instability in the cubic potential was analyzed much earlier, see \cite{Hirsch,Horsthemke,Hanggi} and references therein.)
We will compare our results with those of Refs.~\cite{Ornigotti,SilerPRL,Ryabov} as we move along.

Equations~(\ref{SDE-n}) and (\ref{SDE-nd}), with integer $n>1$, will be in the focus of our attention. Before we proceed, however, let us establish some scaling properties of the problem and slightly simplify the notation (see also Ref. \cite{Ornigotti}). The scale invariance of the power-law potential $U(r)$  makes it possible, in arbitrary spatial dimension, to get rid of the parameters $g$ and $D$. Indeed, rescaling time by the intrinsic time $\tau =  g^{-\frac{2}{n+1}} D^{-\frac{n-1}{n+1}}$ and the coordinate by the characteristic diffusion length $\ell =(D\tau)^{1/2} =(D/g)^{\frac{1}{n+1}}$, one can bring Eq.~(\ref{SDE-n}) to a parameter-free form. We thus can set $g=D=1$ and restore the constants $g$ and $D$ in some of the final results.

We will often  assume for simplicity that the particle starts at the origin. In this case, the average blowup time $\langle T \rangle$ and the probability distribution $\mathcal{P} (T)$ of the blowup time can be written, in arbitrary spatial dimensions, as
\begin{equation}
\label{Tscaling}
\langle T \rangle= \alpha_n \tau = \alpha_n g^{-\frac{2}{n+1}} D^{-\frac{n-1}{n+1}}
\end{equation}
and
\begin{equation}
\label{exactscaling}
\mathcal{P} (T) = \frac{1}{\tau} F_n\left(\frac{T}{\tau}\right)\,,
\end{equation}
respectively. Below we determine the dimensionless coefficient $\alpha_n$ which depends on $n$ and $d$. The dimensionless scaling function $F_n(z)$ depends on the rescaled time $z=T/\tau$, $n$ and $d$ and satisfies the normalization condition $\int_0^\infty dz\,F_n(z)=1$. Below we determine several characteristics of the scaling function $F_n(z)$. In particular, we compute the tails of $F_n(z)$ and show that their scaling behavior is
\begin{equation}
\label{tailsofF}
- \ln F_n(z) \sim
\begin{cases}
z\,,                                   & z \to \infty\,,\\
z^{-\frac{n+1}{n-1}}\,,       & z \to 0\,,
\end{cases}
\end{equation}
in any spatial dimension.

In Secs.~\ref{theory}--\ref{distr} we consider the one-dimensional problem as described by Eq.~(\ref{SDE-n}). The blowup in higher dimensions is considered in Sec.~\ref{sec:high}. Section \ref{conclusions} presents our conclusions and suggests possible directions for further work. The cumulants for the blowup time of a particle in a quartic potential starting at the origin are presented in Appendix \ref{ap:cum}. In Appendix \ref{ap:nozero} we show that all the eigenvalues of a Fokker-Planck  operator which appears in our calculations of $\mathcal{P}(T)$ are positive.

\section{Average blowup time}
\label{theory}

To determine the average blowup time $\Theta(x)$ for the particle starting at the initial position $x$ we employ the {\em backward} Fokker-Planck equation, see, \textit{e.g.}, \cite{SR}. An advantage of this approach is that it circumvents the description of the preceding particle dynamics. The backward Fokker-Planck, corresponding to the Langevin equation~(\ref{SDE-n}), is the following linear ordinary differential equation (ODE)
\begin{equation}
\label{avtimeeq}
\Theta^{\prime\prime}(x)+ x^n \Theta^{\prime}(x) = -1.
\end{equation}
The escape to $x=+\infty$ is always feasible, and it is described by the absorbing boundary condition there:
\begin{equation}
\label{BCinfinity}
\Theta(x\to+\infty) = 0\,.
\end{equation}
For odd $n$, the function $\Theta(x)$ is an even function of $x$. In this case we can use the condition
\begin{equation}
\label{BCzero}
\Theta^{\prime}(x=0) = 0
\end{equation}
as the second boundary condition, and only consider the region of $x>0$. Solving Eq.~\eqref{avtimeeq} subject to the boundary conditions \eqref{BCinfinity}--\eqref{BCzero} yields
\begin{equation}
\label{T-int-m}
\Theta(x) = \int_x^\infty dz \int_0^z dy\,\exp\!\left(\frac{y^{n+1}-z^{n+1}}{n+1}\right).
\end{equation}

The average blowup time $\Theta(x)$ decays algebraically when $x\to \infty$. The subleading correction is also algebraic,
\begin{equation}
\label{av-asymp}
\Theta(x) = \frac{1}{(n-1)x^{n-1}}+\frac{1}{2\,x^{2n}}+\ldots,
\end{equation}
which is most easily derived directly from Eq.~\eqref{avtimeeq}. The leading term is just the deterministic blowup time, \textit{cf.} Eq.~\eqref{determ}.

\begin{figure}
\centering
\includegraphics[width=7.89cm]{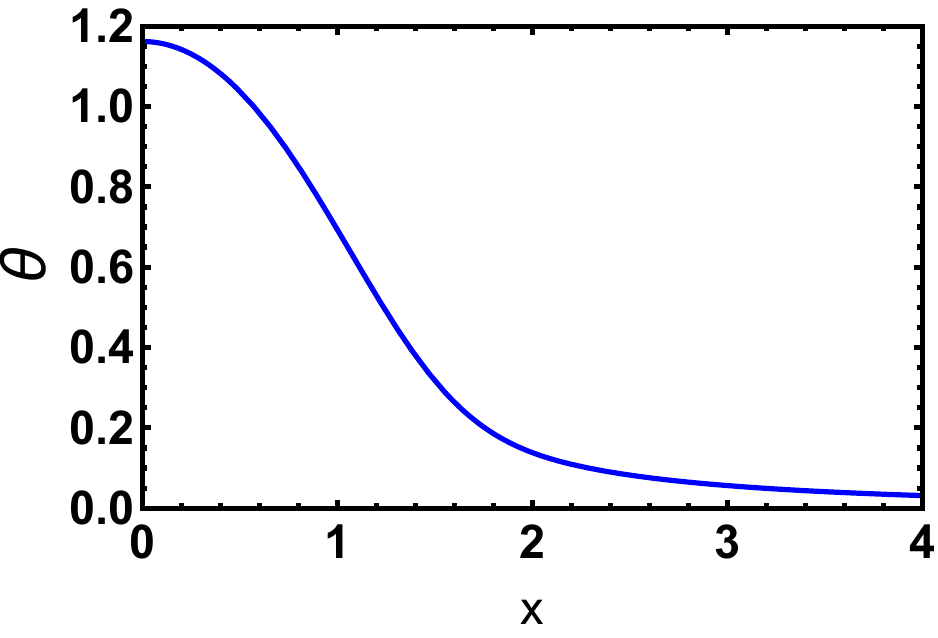}
\caption{The average blowup time $\Theta(x)$ versus the initial position $x$ for $n=3$.  The analytical solution is given by Eq.~\eqref{T-sol}.}
\label{Fig:T-1-4}
\end{figure}

For $n=3$, the  integral representation \eqref{T-int-m} can be expressed through the hypergeometric function with four indexes, and we obtain
\begin{equation}
\label{T-sol}
\Theta(x) = \frac{\pi\,\Gamma\big(\frac{5}{4}\big)}{2\,\Gamma\big(\frac{3}{4}\big)}
-\frac{x^2}{2}\,F\!\left(\frac{1}{2},1;\frac{5}{4},\frac{3}{2}; -\frac{x^4}{4}\right),
\end{equation}
see Fig.~\ref{Fig:T-1-4}.

Specializing \eqref{T-int-m} to $x=0$ we express $\langle T\rangle = \Theta(0)$ through gamma functions
\begin{equation}
\label{avtimeeven}
 \langle T\rangle = C_n, \; C_n\equiv \frac{\Gamma\!\left(\frac{1}{n+1}\right) \Gamma\!\left(\frac{2}{n+1}\right) \Gamma\!\left[\frac{n-1}{2(n+1)}\right]}{4^{\frac{1}{n+1}} \sqrt{\pi} (n+1)^{\frac{2n}{n+1}}}
\end{equation}
for odd $n\geq 3$. [For $n=1$, the escape time is infinite, so the prediction of Eq.~\eqref{avtimeeven} remains correct in this case.] In particular, for $n=3$ we obtain
\begin{equation}
  \langle T\rangle \big|_{n=3}= \sqrt{\frac{2}{gD}}\,\left[\Gamma\left(\frac{5}{4}\right)\right]^2,
\end{equation}
where we have restored $g$ and $D$.

For even $n$, the average blowup time $\Theta(x)$ is not symmetric with respect to the origin, and we must consider the whole interval $|x|<\infty$. The symmetry condition (\ref{BCzero}) should be replaced by the boundary condition \cite{SR}
\begin{equation}
\label{BCminusinfinity}
\Theta^{\prime}(x\to -\infty) = 0.
\end{equation}
(This is of course obeyed for odd $n$ as well.)

Solving Eq.~\eqref{avtimeeq} subject to the boundary conditions \eqref{BCinfinity} and \eqref{BCminusinfinity} yields
\begin{equation}
\label{T-int-even}
\Theta(x) = \int_x^\infty dz \int_{-\infty}^z dy\,\exp\!\left(\frac{y^{n+1}-z^{n+1}}{n+1}\right).
\end{equation}
For even $n\geq 2$ this gives
\begin{equation}
\label{av-even}
 \langle T\rangle = \Theta(0) = (n+1)^\frac{2}{n+1} \left[\Gamma\left(1+\frac{1}{n+1}\right)\right]^2+C_n
\end{equation}
with $C_n$ defined in Eq.~\eqref{avtimeeven}. For even $n\geq 2$ the average blowup time to $+\infty$ is finite even if the particle starts at $x= -\infty$. The ratio of the blowup time $\Theta(-\infty)$ to $\langle T\rangle=\Theta(0)$ is
\begin{equation}
\label{ratio}
 \frac{\Theta(-\infty)}{\Theta(0)} = 1+\frac{1}{1+2\cos\frac{\pi}{n+1}}
\end{equation}
This ratio decreases from $3/2$ at $n=2$ to $4/3$ at $n\to\infty$.

\begin{figure}
\centering
\includegraphics[width=7.8cm]{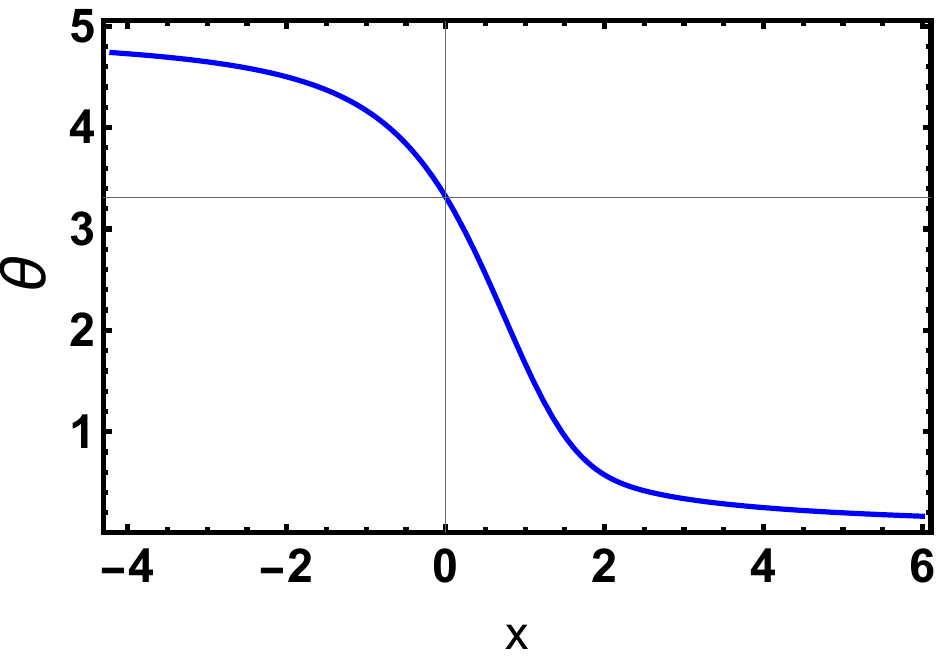}
\caption{The average blowup time $\Theta(x)$ versus the initial position $x$ for $n=2$, see Eq.~(\ref{T-int-even}). Here $\Theta(-\infty)/\Theta(0)=3/2$, as predicted by Eq.~(\ref{ratio}).}
\label{Fig:Tav2}
\end{figure}

Figure \ref{Fig:Tav2} shows a plot of the function $\Theta(x)$, as defined by Eq.~(\ref{T-int-even}), in the particular case $n=2$. Note the asymmetric form of $\Theta(x)$ and its monotone behavior as a function of $x$. If the particle starts from the origin, we obtain
\begin{equation}
  \langle T\rangle \big|_{n=2}= \frac{2 \cdot 3^{2/3} \left[\Gamma \left(4/3\right)\right]^2} {g^{1/3}D^{2/3}}
\end{equation}
where we have restored $g$ and $D$.

\section{Probability distribution of blowup time}
\label{distr}

We will use two different methods for determining the probability distribution of blowup time $\mathcal{P} (T)$ when starting from $x=0$.  The first method is based on solving a linear ODE the one obtains for the Laplace transform of $\mathcal{P} (T)$. The inverse Laplace transform then gives the distribution itself.  The second method is based on the determination of the probability distribution $P(x,t)$ of the particle position at time $t$ by solving the (forward) Fokker-Planck equation, corresponding to the Langevin equation~(\ref{SDE-n}). Then one obtains $\mathcal{P} (T)$ by calculating the probability current at $x\to +\infty$. The two methods turn out to be complementary in this problem,  as we will see shortly.

\subsection{$\mathcal{P}(T)$ via Laplace transform}
\label{Laplace}

\begin{figure}
\includegraphics[width=0.46\textwidth,clip=]{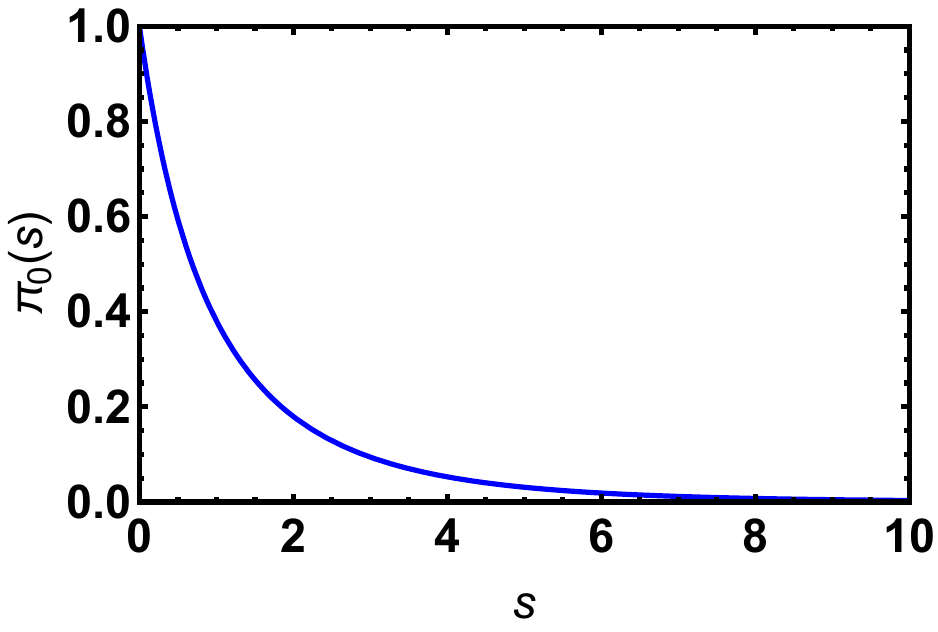}
\caption{The Laplace transform of $\mathcal{P}(T)$ for $n=3$, as described by Eq.~(\ref{Pi-0}).}
\label{LTn3}
\end{figure}

Here we denote the initial position of the particle by $x$. Let us consider the probability distribution $\mathfrak{p}(T,x)$ of the blowup time $T$ for this initial condition. The Laplace transform of this distribution,
\begin{equation}
\label{LT}
\Pi(s,x)=\left\langle e^{-s T(x)}\right\rangle = \int_0^\infty dT\,e^{-sT} \mathfrak{p}(T,x)\,,
\end{equation}
is described by another linear ODE
\begin{equation}
\label{Pi-eq}
\Pi^{\prime\prime}(x) +x^n\,\Pi^{\prime}(x) = s \Pi(x)\,,
\end{equation}
subject to the boundary conditions
\begin{equation}
\label{BCslaplace}
\Pi(x\to+\infty)=1 \quad \text{and} \quad \Pi^{\prime}(x\to -\infty)=0\,,
\end{equation}
see, \textit{e.g.}, Refs.~\cite{SR,KRB,KR18}.  We succeeded in solving Eq.~\eqref{Pi-eq} in explicit form only for $n=3$. In this case, and generally for odd $n$, the Laplace transform is an even function, $\Pi(s,-x)=\Pi(s,x)$, and therefore the second condition in Eq.~(\ref{BCslaplace}) can be replaced by
\begin{equation}
\Pi^{\prime}(x=0) = 0\,.
\end{equation}
The solution reads
\begin{equation}
\label{Pi-sol}
\Pi(s,x)=\frac{\text{HeunB}[s/4, 0, 1/2, 0, 1/2; x^2]}{\text{HeunB}[s/4, 0, 1/2, 0, 1/2; \infty]}.
\end{equation}
where $\text{HeunB}[a_1,a_2,a_3,a_4,a_5; z]$ denotes the bi-confluent Heun function \cite{NIST}.
When the particle starts at the origin, $x=0$, the Laplace transform \eqref{Pi-sol} simplifies to
\begin{equation}
\label{Pi-0}
\Pi_0(s)=\frac{1}{\text{HeunB}[s/4, 0, 1/2, 0, 1/2; \infty]}
\end{equation}
by virtue of the identity $\text{HeunB}[a_1,a_2,a_3,a_4,a_5; 0] = 1$, a normalization condition for the Heun function, which holds for arbitrary values of its five parameters. A plot of $\Pi_0(s)$ is shown in Fig.~\ref{LTn3}.

\begin{figure}
\includegraphics[width=0.46\textwidth,clip=]{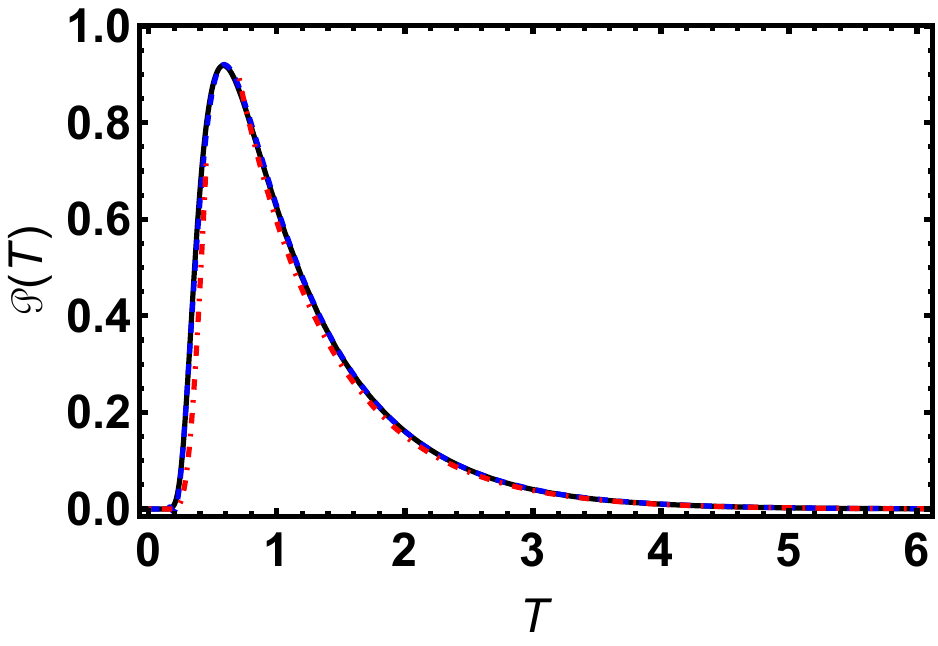}
\caption{The blowup time distribution $\mathcal{P}(T)$ in the rescaled variables, obtained from numerical inverse Laplace transform of Eq.~(\ref{Pi-0}) (blue dashed line) and from numerical solution of the Fokker-Planck equation (solid black line) for $n=3$. These two lines are almost indistinguishable. Also shown, by the two red dash-dotted lines, are the long-time asymptotic~(\ref{longTn}) and the short-time asymptotic~(\ref{smallT3}).}
\label{invL}
\end{figure}

The inverse Laplace transform,
\begin{equation}
\label{inversetr}
 \mathcal{P}(T) = \frac{1}{2\pi \ii}\int_{\gamma-\ii \infty}^{\gamma+\ii \infty} e^{sT} \Pi_0(s) \,ds,
\end{equation}
with $\Pi_0(s)$ given by \eqref{Pi-0} yields the blowup time distribution for $n=3$. The dashed blue line in Fig.~\ref{invL} shows the resulting distribution $\mathcal{P}(T)$ obtained by performing the inverse Laplace transform numerically.

As one can check (again, numerically), the closest to the origin singularity of $\Pi_0(s)$, given by Eq.~\eqref{Pi-0}, is a simple pole:
\begin{equation}
\label{pole}
\Pi_0(s)=\frac{A}{s+a}+\ldots,
\end{equation}
with $a= 1.36859\ldots$ and $A\simeq 2.5$.  The singular behavior \eqref{pole} implies that the long-time tail of the blowup time distribution is purely exponential:
\begin{equation}
\label{large-3}
\mathcal{P}(T\to \infty)\simeq A e^{-aT}\,.
\end{equation}
This long-time tail is also shown, by a dashed red line, in Fig.~\ref{invL}. In the next subsection we will establish an intimate connection of this tail with the slowest decaying mode of the pertinent Fokker-Planck operator. With the units restored, the long-time tail \eqref{large-3} reads
\begin{equation}
\label{longTn}
\mathcal{P} (T\to \infty) \simeq A \exp(-a g^{\frac{2}{n+1}} D^{\frac{n-1}{n+1}} T)\,.
\end{equation}

Figure~\ref{invL} also suggests that $\mathcal{P}(T)$ exhibits an essential singularity at $T=0$. As we will see shortly, this is indeed an essential singularity. It is controlled by the $s\to \infty$ asymptotic of the bi-confluent Heun function appearing in Eq.~(\ref{Pi-0}) which, unfortunately, does not seem to be available. In subsection \ref{sub:0}, we will determine the leading-order $\mathcal{P}(T\to 0)$ asymptotic and characterize the essential singularity by using the optimal fluctuation method. In this way, we will also obtain the optimal (that is, the most likely) trajectory $x(t)$ conditioned on an unusually fast blowup.

The Laplace transform encodes the average, the variance, and all higher cumulants of the blowup time. In Appendix \ref{ap:cum}, we present analytical expressions for the cumulants of the blowup time for the particle in the quartic potential starting at the origin.

\subsection{$\mathcal{P}(T)$ via Fokker-Planck equation}
\label{sec:FPE}

Here we denote by $x$ the position of the particle at time $t$. The probability distribution $P(x,t)$ is described by the Fokker-Planck equation, corresponding to the Langevin equation~(\ref{SDE-n}). In the rescaled variables $g=D=1$  we have
\begin{equation}
\label{FPeq}
\frac{\partial P}{\partial t}= -\frac{\partial}{\partial x} \left(x^n P\right)+ \frac{\partial^2 P}{\partial x^2}.
\end{equation}
We assume that the particle starts at the origin, so that the initial condition for Eq.~(\ref{FPeq}) is
\begin{equation}
\label{incondP}
P(x,t=0) = \delta (x)\,.
\end{equation}
The solution can be sought via expansion over the eigenfunctions $u_k(x)$, $k=1,2,\ldots$, of the Fokker-Planck operator. The eigenfunctions $u_k(x)$ obey the linear second-order ODE
\begin{equation}\label{ODEu}
u_k^{\prime\prime}(x) - \left[x^n u_k(x)\right]^{\prime}+ \lambda_k u_k(x) = 0\,,
\end{equation}
where $\lambda_k$ are the eigenvalues. The boundary conditions are $u(|x|\to \infty) = 0$.

To determine the spectrum of eigenvalues $\lambda$, let us first transform $u_k$ to a new variable
\begin{equation}
\label{transform1}
v_k(x)=u_k(x) \exp\left[-\frac{x^{n+1}}{2(n+1)}\right]\,.
\end{equation}
(The same standard transformation was applied to Eq.~(\ref{ODEu}) for $n=2$ in Ref. \cite{Ryabov}.)
Equation (\ref{ODEu}) becomes a Schr\"{o}dinger equation
\begin{equation}
\label{Schroedinger}
v_k^{\prime\prime}(x)+\left[\lambda_k -V(x,n)\right] v_k(x) =0
\end{equation}
for an effective quantum particle with energy $\lambda_k$ in the potential
\begin{equation}\label{V(x,n)}
V(x,n)=\frac{x^{2 n}}{4}+\frac{n x^{n-1}}{2}\,.
\end{equation}
For $n>1$ the potential $V(x,n)$ is confining, hence the spectrum $\lambda_k$, $k=1,2, \dots$ is discrete. Furthermore, for odd $n$, we have $V(x,n)>0$ for all $x$; therefore, all the eigenvalues $\lambda_k$ are strictly positive. For even $n$ the potential is negative on the interval $-(2n)^{1/(n+1)}<x<0$. However, as we show in Appendix \ref{ap:nozero}, all the eigenvalues $\lambda_k$ are still positive, as to be expected.

Therefore, the probability distribution of the particle position at time $t$ has the form \begin{equation}\label{expansion}
P(x,t) = \sum_{k=1}^{\infty} b_k u_k (x) e^{-\lambda_k t}\,,
\end{equation}
where we have omitted the index $n$ for brevity.  The expansion amplitudes $b_k$  can be found by transforming Eq.~(\ref{ODEu}) into the self-adjoint form
\begin{equation}\label{selfadjointgeneral}
\frac{d}{dx}\left(e^{-\frac{x^{n+1}}{n+1}} \frac{du}{dx}\right) +\left(\lambda -n x^{n-1}\right) e^{-\frac{x^{n+1}}{n+1}} u=0\,,
\end{equation}
and projecting the delta-function, see Eq.~(\ref{incondP}), onto the eigenfunctions $u_k(x)$:
\begin{eqnarray}
  b_k&=& \frac{\int_{-\infty}^{\infty} dx\, \delta(x) u_k(x) e^{-\frac{x^{n+1}}{n+1}}}{\int_{-\infty}^{\infty} dx\,u_k^2(x) e^{-\frac{x^{n+1}}{n+1}}}  \nonumber\\
&=&  \left[\int_{-\infty}^{\infty} dx\,u_k^2(x) e^{-\frac{x^{n+1}}{n+1}}\right]^{-1}\,, \label{bk}
\end{eqnarray}
where (without loss of generality) we set the normalization $u_k(0)=1$. Note that the general expansion~(\ref{expansion}) remains valid for arbitrary initial condition. Only the amplitudes $b_k$ depend on the initial condition.

Once $P(x,t)$ is found, the probability distribution $\mathcal{P}(T)$ of the blowup time can be determined by the probability flux to infinity:
\begin{eqnarray}
\label{relation}
  \mathcal{P}(T) &=& \mu_n \lim_{x\to \infty}\left[-D \frac{\partial P(x,T)}{\partial x}+x^n P(x,T)\right] \nonumber \\
  &=& \mu_n \lim_{x\to \infty}x^n P(x,T)\,.
\end{eqnarray}
The last equality is correct because of the asymptotic behavior of the eigenfunctions $u_k(x) \sim x^{-n}$ at $x\to \infty$. As a result, the first term on the right-hand side in the first line of Eq.~(\ref{relation}) vanishes in the limit of $x\to \infty$, whereas the second term yields an expression that depends only on $T$ as it should. Finally,  the factor $\mu_n$ comes from the normalization of the distribution $\mathcal{P}(T)$ to $1$:
\begin{equation}
\label{normalization}
\int_0^T \mathcal{P}(T)\,dT = 1.
\end{equation}
We have $\mu_n=1$ for even $n$, where the escape is only to $+\infty$, and $\mu_n=2$ for odd $n$, where the escape is possible to $-\infty$ as well.

\subsubsection{$T\to \infty$}
\label{sub:inf}

The long-time limit corresponds to the strong inequality $T\gg \langle T \rangle$. By virtue of Eq.~(\ref{expansion}), the long-time behavior of $P(x,t)$,
\begin{equation}
\label{longtimeansatz}
P(x, t \to \infty) \simeq  b_1 e^{-\lambda_1 t} u_1(x),
\end{equation}
is determined by the minimum eigenvalue $\lambda_1$ and the ground state eigenfunction $u_1(x)$.

\begin{figure}
\includegraphics[width=0.44\textwidth,clip=]{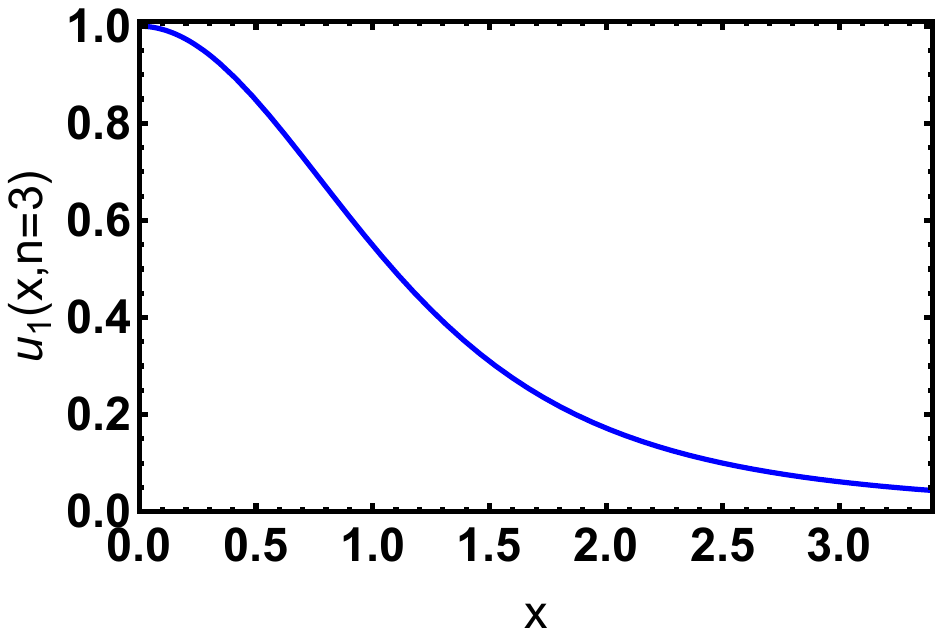}
\caption{The eigenfunction $u_1(x)$, corresponding to the minimum eigenvalue $\lambda_1$, see Eqs.~(\ref{ODEu}) and~(\ref{longtimeansatz}), for $n=3$.  $u_1(x)$ is arbitrarily normalized by setting $u_1(0)=1$. The eigenfunction is even, $u_1(x) = u_1(-x)$, and shown only in the $x>0$ range.}
\label{eigenfunction3}
\end{figure}

We now provide more details for $n=3$ and $n=2$. When $n=3$, all the modes $u_k(x)$ are symmetric with respect to the origin, and we obtain
\begin{equation}\label{ukgeneral}
u_k(x) = C_k \,\text{HeunB}\!\left[-\tfrac{\lambda_k}{4},-\tfrac{3}{4},\tfrac{1}{2},0,-\tfrac{1}{2}; x^2\right]\,,
\end{equation}
with amplitudes $C_k$ depending on the initial condition. The eigenvalues are determined from the boundary condition $u(|x|\to \infty) = 0$. The minimum eigenvalue is $\lambda=\lambda_1 = 1.36859\dots$, in perfect agreement with our Laplace-transform result in Eq.~(\ref{pole}). The ground state eigenfunction $u_1(x)$ is depicted in Fig.~\ref{eigenfunction3}. At large $x$, it falls off as $x^{-3}$, like all the other modes for $n=3$.

The amplitude $b_1$ is given by Eqs.~(\ref{bk}) and (\ref{ukgeneral}) for $k=1$, and we obtain
\begin{equation}
  b_1 =  \left[\int_{-\infty}^{\infty} dx\,u_1^2(x) e^{-\frac{x^4}{4}}\right]^{-1} =0.680724 \dots\,. \label{a1}
\end{equation}
Now we turn to Eq.~(\ref{relation}) and evaluate numerically the limit $\lim_{x\to \infty} x^3 u_1(x) \simeq 1.86$.  The numerical extrapolation to $x=+\infty$ is conveniently done by exploiting our knowledge of the $x^{-5}$ subleading correction to the $u_k(x\to +\infty) \sim x^{-3}$ leading-order behavior. Restoring the units, we reproduce Eq.~(\ref{longTn}).

\begin{figure}
\includegraphics[width=0.44\textwidth,clip=]{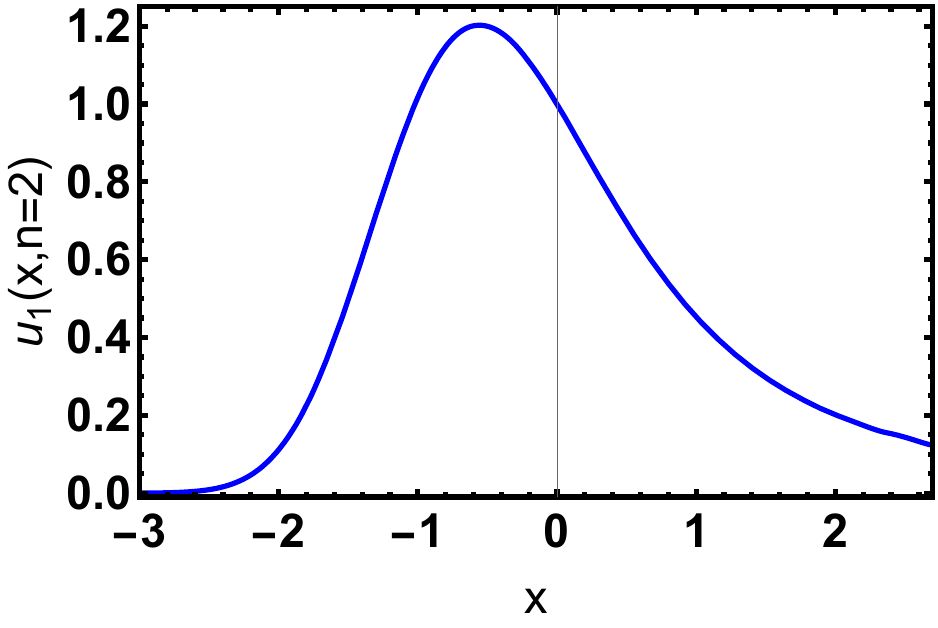}
\caption{The numerically found eigenfunction $u_1(x)$, corresponding to the minimum eigenvalue $\lambda_1$, see Eqs.~(\ref{ODEu}) and (\ref{longtimeansatz}), for $n=2$. The eigenfunction $u_1(x)$ is arbitrarily normalized by setting $u_1(0)=1$.}
\label{eigenfunction2}
\end{figure}

Now we turn to $n=2$. Here we determined the minimum eigenvalue, $\lambda_1\simeq 0.354$, and the ground state eigenfunction, which is shown in Fig.~\ref{eigenfunction2}, numerically. As expected, this eigenfunction is asymmetric with respect to the origin, and its maximum is at a negative $x$, that is it is shifted against the deterministic force. Further, $u_1(x)$ decays faster than exponentially at $x\to -\infty$ and only algebraically, as $x^{-2}$, at $x\to \infty$. These remarkable behaviors, viz., the shift of the maximum and the strong asymmetry of the tails of $u_1(x)$ for $n=2$, were predicted in Ref.~\cite{Ornigotti} and observed in experiment \cite{SilerPRL}. Furthermore, the long-time asymptotic of the position distribution $P(x,t)$ as described by Eq.~(\ref{longtimeansatz}) for $n=2$, coincides up to a normalization factor with the ``quasistationary distribution" which was the focus of attention of Refs. \cite{Ornigotti,SilerPRL}.

The amplitude $b_1$ is again given by Eqs.~(\ref{bk}) and (\ref{ukgeneral}) for $k=1$:
\begin{equation}
  b_1 =\left[\int_{-\infty}^{\infty} dx\,u_1^2(x) e^{-\frac{x^3}{3}}\right]^{-1} \simeq 0.36 \,, \label{a12}
\end{equation}
where we have used the numerically found eigenfunction $u_1(x)$. Now we turn to Eq.~(\ref{relation}) and evaluate numerically $\lim_{x\to \infty} x^2 u_1(x) \simeq  1.22$. Here the numerical extrapolation to $x=+\infty$ exploits an $x^{-3}$ subleading correction to the $u_k(x\to +\infty) \sim x^{-3}$ leading-order behavior. Summarizing and restoring the units, we arrive at 
\begin{equation}
\label{longTn=2}
\mathcal{P} (T\to \infty) \simeq 0.44\,g^\frac{2}{3} D^\frac{1}{3} e^{-0.35 \,g^\frac{2}{3} D^\frac{1}{3}T},\;n=2.
\end{equation}
Equation (\ref{longTn=2}) holds, up to numerical coefficient, for any initial position $x$ of the particle. The analytical determination of the $x-$dependent factor appears challenging.  (When $n=3$, the $x-$dependent factor is easy to extract from the Laplace transform \eqref{Pi-sol}. Below, we demonstrate how to do it in the general case of arbitrary spatial dimension, see \eqref{detailed}.)


To conclude this subsection, we provide a qualitative explanation for the fact that, for all $n>1$, the long-time tail of  $\mathcal{P}(T)$ decays exponentially. To avoid blowup for an unusually long time, the particle must stay sufficiently close to the origin, where the deterministic repulsion force is small. In a similar classical problem of survival of a Brownian particle on an interval $|x|<\ell_0$ against absorption by the edges of the interval \cite{SR,KRB}, the survival probability of the particle also decays exponentially, with the decay rate of order $D/\ell_0^2$. Similarly to the blowup problem, this exponential decay corresponds to the ground state of the corresponding operator (which, in that case, is a pure diffusion). Adapting our blowup problem to this simple model problem, we recall that the only intrinsic length scale in the blowup problem is $\ell =(D/g)^{\frac{1}{n+1}}$. Setting $\ell_0=\ell$, we observe that the ensuing decay rate matches that of Eq.~(\ref{longTn}) up to a numerical factor not accounted for by this simple order-of-magnitude estimate.

\subsubsection{$T\to 0$}
\label{sub:0}

The short-time limit corresponds to $T\ll \langle T \rangle$. All the eigenfunctions in Eq.~(\ref{expansion}) now contribute to the solution, and it is much more efficient to use the optimal fluctuation method (OFM), also known as the weak noise theory. The OFM also gives \emph{the optimal path}, viz., the most likely trajectory $x(t)$ conditioned on an unusually fast blowup and the most likely realization of the noise $\eta(t)$.

\begin{figure}
\includegraphics[width=0.44\textwidth,clip=]{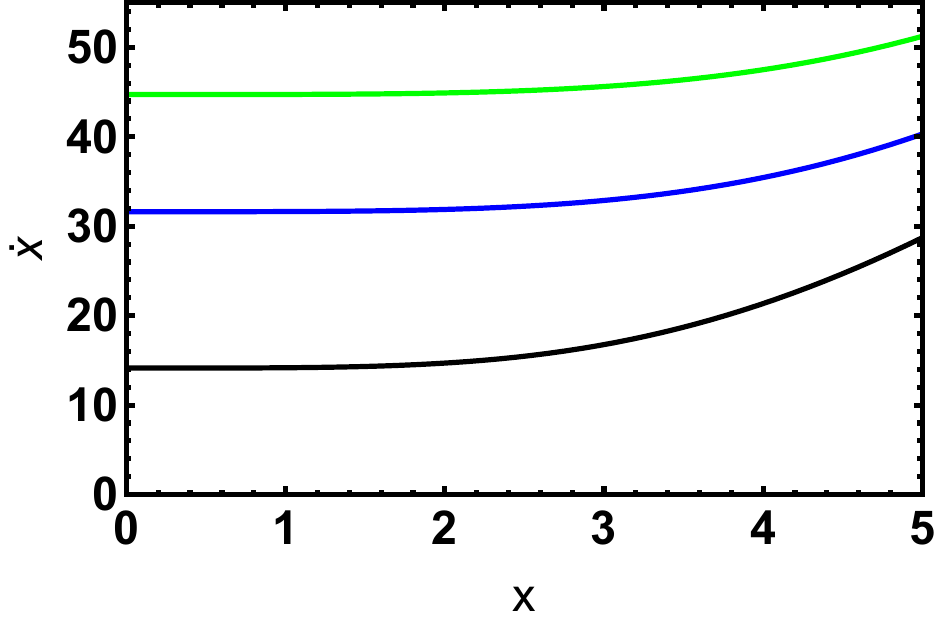}
\caption{Phase plane of the optimal paths to blowup for $n=2$. Shown is $\dot{x}$ vs. $x$ for $E=10^2$ (black), $5\cdot 10^2$ (blue) and $10^3$ (green). The corresponding blowup times, as described by Eq.~(\ref{energy}), are $0.4930\ldots$, $0.3297\dots$ and $0.2772\ldots$, respectively.}
\label{phaseportrait}
\end{figure}

One way of applying the OFM relies on solving the Fokker-Planck equation (\ref{FPeq}) by the exponential ansatz of the WKB type, $P(x,t) = a(x,t) \exp[-S(x,t)/D]$ in the limit of $D\to 0$. In the leading order, one arrives at the Hamilton-Jacobi equation for $S(x,t)$. Once $S(x,t)$ is found, the pre-exponential factor $a(x,t)$ can be determined in the subleading order. Here, we will confine ourselves to the leading-order asymptotic, which ignores the prefactor $a(x,t)$, and use a more direct version of the OFM which applies directly to the Langevin equation~(\ref{SDE-n}). The key idea is that the probability of an unusually fast blowup, with $T\ll \langle T \rangle$,  is dominated by a single optimal path which minimizes the action functional, corresponding to the Langevin Eq.~(\ref{SDE-n}). In the rescaled form, the action functional reads
\begin{equation}
\label{action}
  S=\frac{1}{4} \int_0^T (\dot{x}-x^n)^2\,dt\,,
\end{equation}
and the minimization should be performed subject to the boundary conditions in time
\begin{equation}
\label{BCWKB}
 x(t=0) = 0 \quad\text{and}\quad x(t=T)=+\infty\,,
\end{equation}
where we continue to assume for simplicity that the particle starts at the origin.
The Euler-Lagrange equation
\begin{equation}\label{ELeq}
\ddot{x}(t)-n x^{2n-1}(t)=0\,,
\end{equation}
has the ``energy" integral:
\begin{equation}
\label{energyintegral}
\dot{x}^2(t)- x^{2n}(t) =2 E =\text{const}\,,
\end{equation}
where the constant $E$ parametrizes the blowup time $T$. Equation~(\ref{energyintegral}) describes the phase plane $(x,\dot{x})$ of the system. For example, Fig.~\ref{phaseportrait} shows such a phase plane for $n=2$ and $E=10^2$, $5\cdot 10^2$ and $10^3$.

Equation~(\ref{energyintegral}) reduces the problem of determining the optimal path to integration of the first-order equation $\dot{x} = (2E+x^{2n})^{1/2}$. In particular, the time of escape to intinity as a function of $E$ is the following:
\begin{eqnarray}
  T &=& \int_0^{\infty} \frac{dx}{(2E+x^{2n})^{1/2}} \nonumber \\
     &=& \frac{\Gamma\!\left(1+\frac{1}{2 n}\right) \Gamma\!\left(\frac{n-1}{2 n}  \right) (2E)^{\frac{1}{2}
   \left(\frac{1}{n}-1\right)}}{\sqrt{\pi }}\,.
   \label{time}
\end{eqnarray}
This expression is finite only for $n>1$. Inverting it, we obtain $E$ as a function of $T$:
\begin{equation}
\label{energy}
E=\left[\Gamma\!\left(\frac{2n+1}{2n}\right) \Gamma\!\left(\frac{n-1}{2
   n}\right)\right]^{\frac{2 n}{n-1}}\, \frac{\left(\sqrt{\pi}\, T\right)^{-\frac{2
   n}{n-1}}}{2}\,.
\end{equation}
Now we can evaluate the action $S$ in Eq.~(\ref{action}). Using Eqs.~(\ref{energyintegral}) and (\ref{energy}),
and going over from integration over $t$ from $0$ to $T$ to integration over $x$ from zero to infinity, we obtain
\begin{eqnarray}
S &=& \frac{1}{4} \int_0^T dt\,\left(\sqrt{2E+x^{2n}}-x^n \right)^2\nonumber\\
  &=& \frac{1}{4}\int_0^{\infty} dx\,\frac{\left(\sqrt{2E+x^{2n}}-x^n \right)^2}{\sqrt{2E+x^{2n}}} \,.
\label{action2}
\end{eqnarray}
Evaluating this integral and restoring the units, we obtain the leading-order asymptotic of the short-time tail:
\begin{eqnarray}
  &-&\ln \mathcal{P} (T\to 0) \simeq S(T)\nonumber \\
  &=&\frac{\Gamma
   \left(\frac{3}{2}-\frac{1}{2
   n}\right) \Gamma\!\left(\frac{1}{2
   n}\right) \left[\Gamma
   \left(1+\frac{1}{2 n}\right) \Gamma\!\left(\frac{n-1}{2
   n}\right)\right]^{\frac{n+1}{n-1}} }{4 (n+1)\pi^{\frac{n}{n-1}} D g^{\frac{2}{n-1}}  T^{\frac{n+1}{n-1}}}\,.
 \label{smallTgeneral}
\end{eqnarray}
The $1/D$ scaling in the right-hand side is the hallmark of the OFM. In this regime  the scaling behavior of $\ln \mathcal{P} (T)$ with $T$, as described by Eq.~(\ref{smallTgeneral}), or by the second line in Eq.~(\ref{tailsofF}), immediately follows from dimensional analysis.

As one can see, the short-time tail (\ref{smallTgeneral}) indeed exhibits an essential singularity,
$$
\mathcal{P}(T\to 0) \sim \exp\left(-\,\text{const} \,T^{-\frac{n+1}{n-1}}\right),
$$
which depends on $n$. It is the strongest for $n=2$, and it becomes milder as $n$ is increased. For $n=2$, $3$ and $4$ we obtain
\begin{eqnarray}
 \!-\!\ln \mathcal{P} (T\!\to \! 0)\big|_{n=2} \!&\simeq& \!\frac{\left[\Gamma\!\left(\frac{1}{4}\right)
   \Gamma\!\left(\frac{5}{4}\right)\right]^4}{12
   \pi ^2 D g^2 T^3}\,, \label{smallT2}\\
 \!-\!\ln \mathcal{P} (T\!\to \! 0)\big|_{n=3}\!&\simeq& \!\frac{\Gamma\!\left(\frac{1}{6}\right)
   \left[\Gamma\!\left(\frac{1}{3}\right)
   \Gamma\!\left(\frac{7}{6}\right)\right]^2
   \Gamma\!\left(\frac{4}{3}\right)}{16
   \pi ^{3/2} D g T^2}\,, \label{smallT3}\\
  \!-\!\ln \mathcal{P} (T\!\to\! 0)\big|_{n=4} \!&\simeq& \!\frac{\Gamma\!\left(\frac{1}{8}\right)
   \left[\Gamma\!\left(\frac{3}{8}\right)
   \Gamma\!\left(\frac{9}{8}\right)\right]^\frac{5}{3}
   \Gamma\!\left(\frac{11}{8}\right)}{20
   \pi ^{4/3} D g^\frac{2}{3} T^\frac{5}{3}}\,. \label{smallT4}
\end{eqnarray}

The leading-order asymptotic (\ref{smallT3}) for $n=3$ is shown in Fig.~\ref{invL}, where we introduced a pre-exponential factor $4$ (essentially, a fitting parameter) which is beyond the leading-order OFM.

\begin{figure}
\includegraphics[width=0.30\textwidth,clip=]{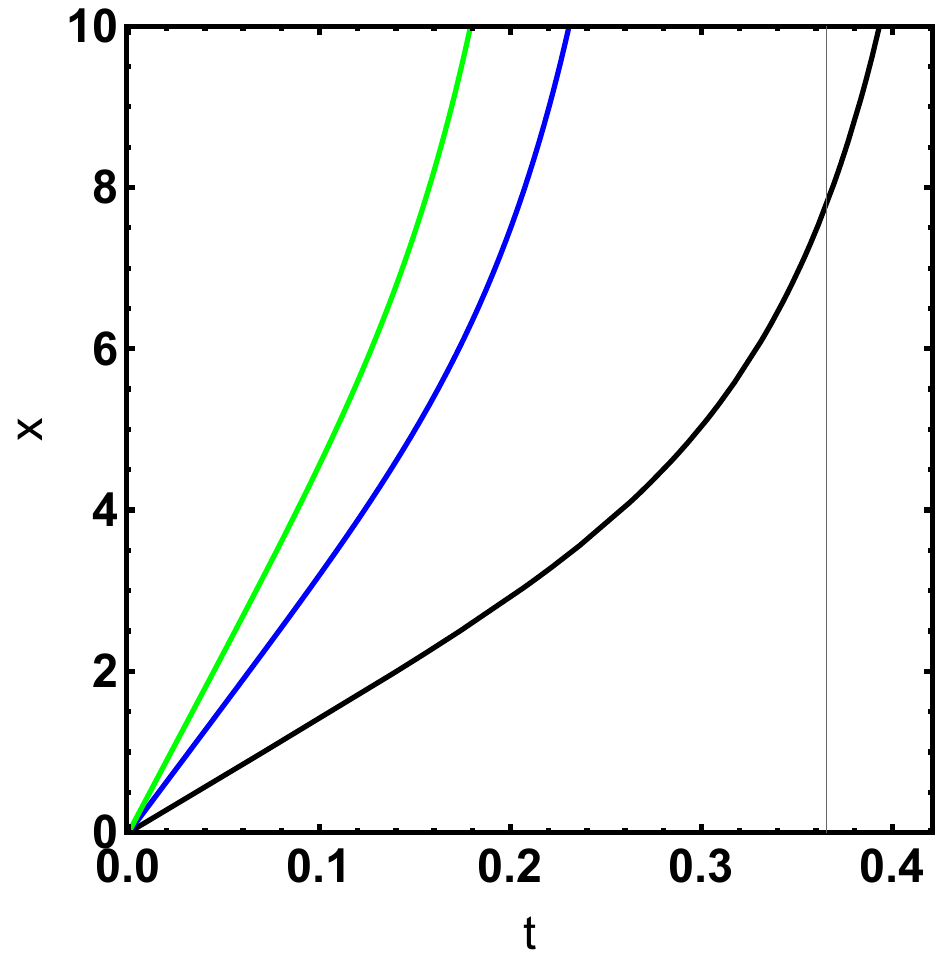}
\includegraphics[width=0.30\textwidth,clip=]{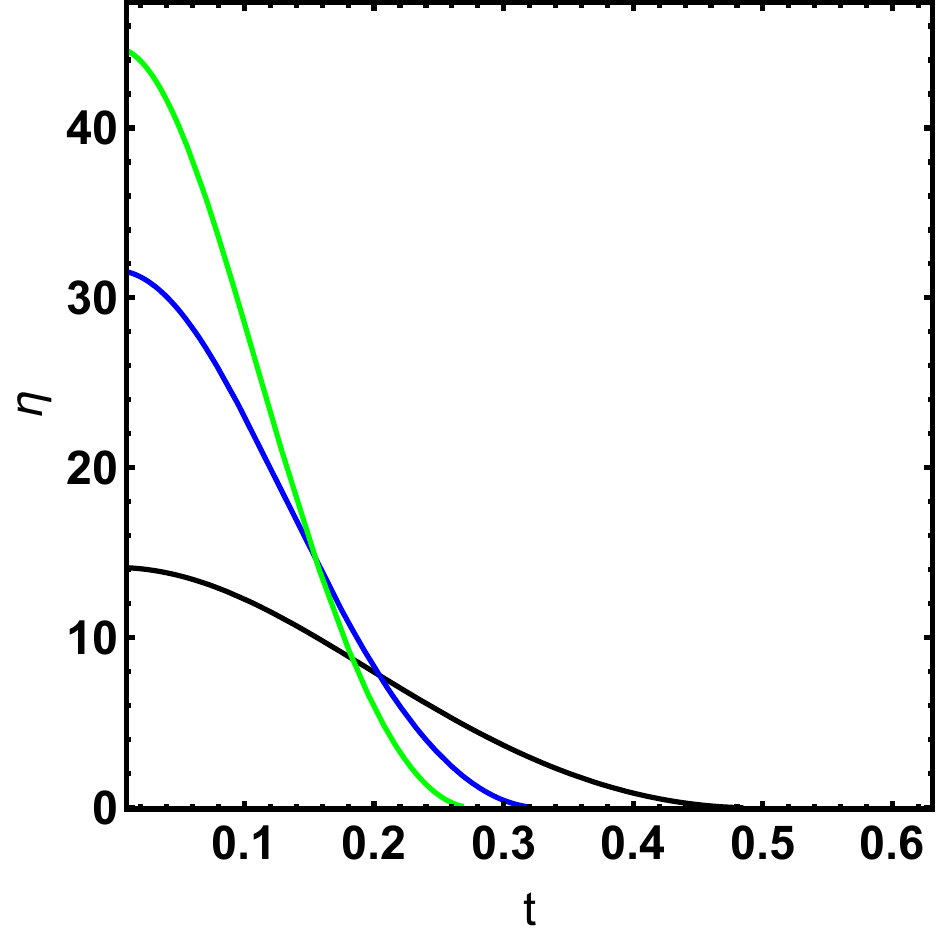}
\caption{The optimal path $x(t)$ (left panel) and  the optimal realization of the noise $\eta(t)$ (right panel) for $n=2$ and $E=10^2$ (black), $5\cdot 10^2$ (blue) and $10^3$ (green).  The corresponding rescaled blowup times are $0.4930\ldots$, $0.3297\dots$ and $0.2772\ldots$, respectively.}
\label{xxivst}
\end{figure}

We obtained Eq.~(\ref{smallTgeneral}) without explicitly calculating the optimal path. The optimal path, however, is interesting in its own right, as it gives insight into the nature of the large deviation in question. Let us establish the optimal path $x(t)$ for $n=2$. Using Eq.~(\ref{energy}), we express the inverse function $t=t(x)$ through the hypergeometric function, and the ``energy" $E=E(T)$ is given by Eq.~(\ref{energy}):
\begin{equation}
\label{t(x)}
t=\int_0^{x} \frac{dy}{(2E+y^4)^{1/2}}=\frac{x}{\sqrt{2E}}\, F\!\left(\frac{1}{4},\frac{1}{2};\frac{5}{4};-\frac{x^4}{2E}\right)\,.
\end{equation}
The left panel of Fig.~\ref{xxivst} shows the optimal paths for three values of energy: $E=10^2, 5\cdot 10^2$ and $10^3$. The right panel of Fig.~\ref{xxivst} shows, for the same values of $E$, the optimal realizations of the noise $\eta(t)$, see Eq.~(\ref{SDE-n}) for $n=2$. These are given in a parametric form by the equation
\begin{equation}
\label{xivsx}
\eta(t) =  \dot{x}(t) - x^n (t) = \sqrt{2E+x^{2n}(t)} - x^n(t)
\end{equation}
and Eqs.~(\ref{energy}) and~(\ref{t(x)}). As one can see, the shorter the blowup time is, the larger is the optimal noise. Also, the optimal noise is the largest in the beginning of the process. Later on it rapidly goes down, the blowup relying more and more on the deterministic force, which ``does the job for free".

\subsubsection{Numerical solution of the Fokker-Planck equation}

\begin{figure}
\includegraphics[width=0.3735\textwidth,clip=]{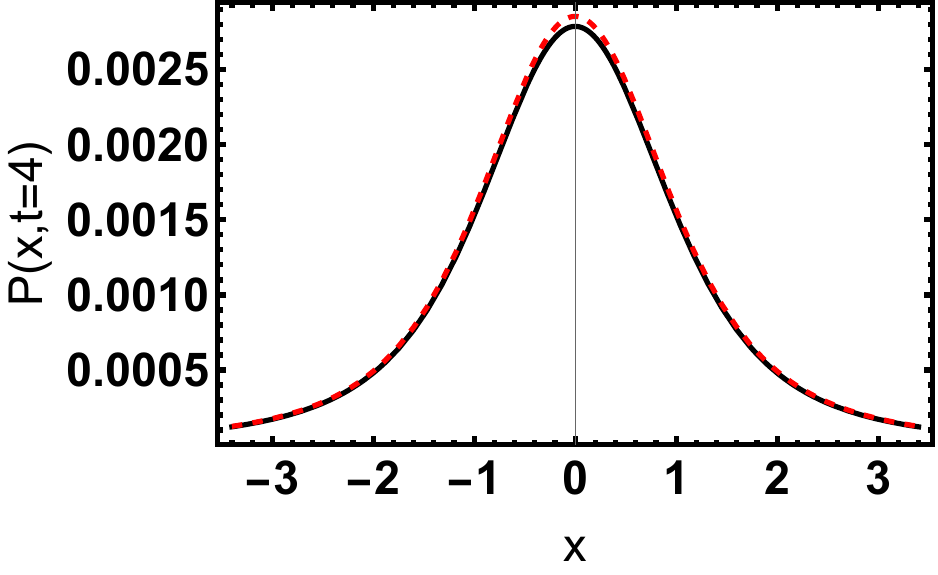}
\includegraphics[width=0.378\textwidth,clip=]{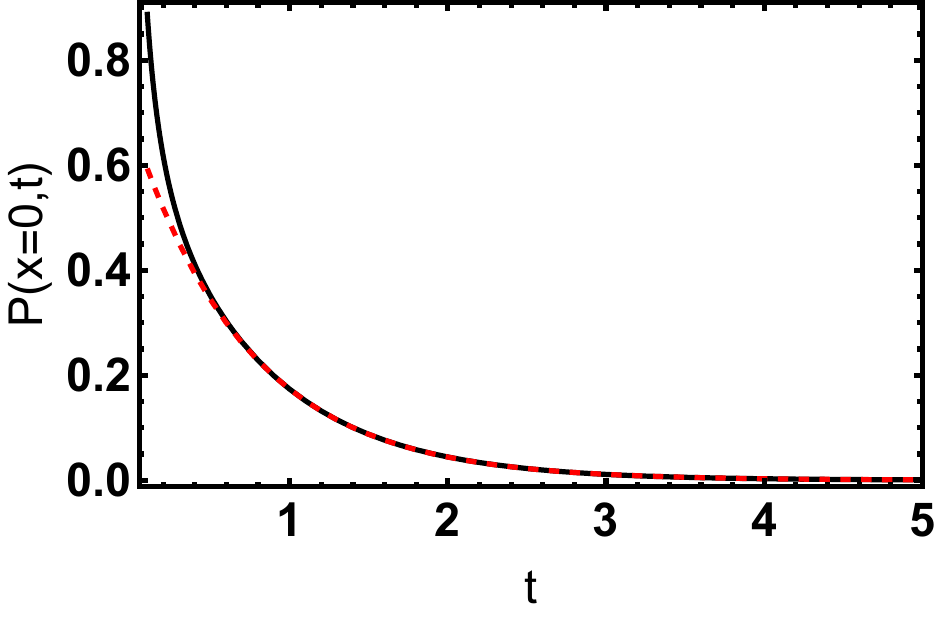}
\caption{The position distribution $P(x,t)$, computed numerically  for $n=3$ (solid lines). Dashed lines: predictions from the single-mode long-time asymptotic~(\ref{longtimeansatz}).}
\label{num1}
\end{figure}

\begin{figure}
\includegraphics[width=0.388\textwidth,clip=]{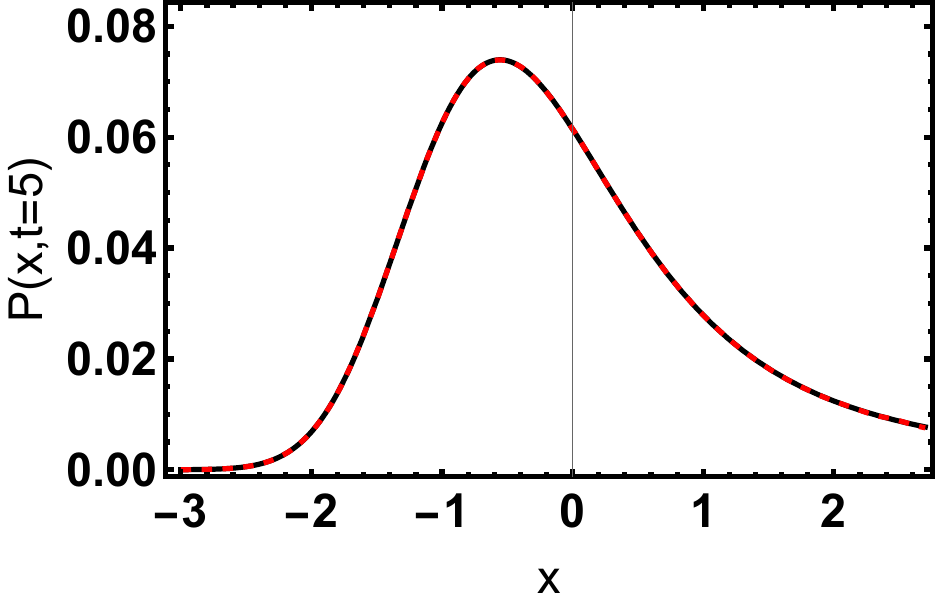}
\includegraphics[width=0.397\textwidth,clip=]{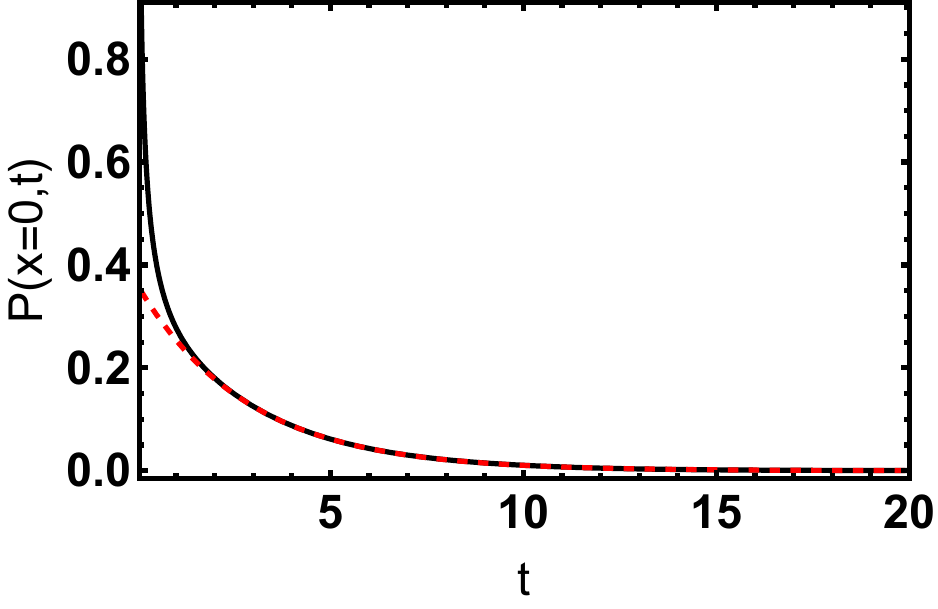}
\caption{The position distribution $P(x,t)$, computed numerically for $n=2$ (solid lines). Dashed lines: predictions from the single-mode long-time asymptotic~(\ref{longtimeansatz}).}
\label{num3}
\end{figure}

We also solved the Fokker-Planck equation (\ref{FPeq}) numerically, with a regularized delta-function at $x=0$ as the initial condition. Figures~\ref{num1} and ~\ref{num3} compare the computed position distributions $P(x,t)$ with the long-time asymptotics  (\ref{longtimeansatz}) and the short-time asymptotics (\ref{smallTgeneral}) for $n=3$ and $2$, respectively.  The solid black line in Fig.~\ref{invL} shows the resulting blowup time distribution $\mathcal{P}(T)$ for $n=3$, which perfectly agrees with that obtained by the numerical inverse Laplace transform in Sec. \ref{Laplace}.

\begin{figure}
\includegraphics[width=0.44\textwidth,clip=]{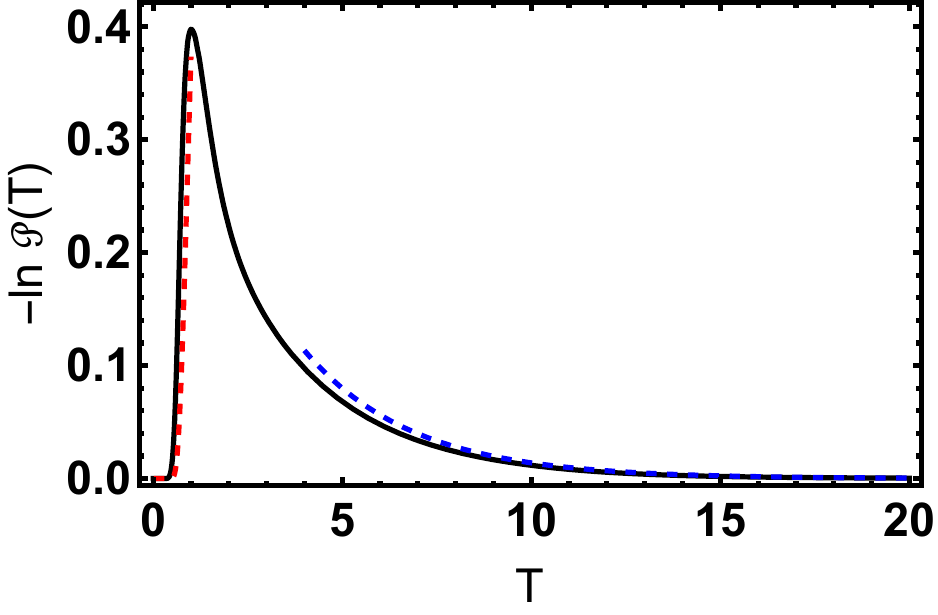}
\caption{Solid lines: the blowup time distribution $\mathcal{P}(T)$ computed from a numerical solution of the Fokker-Planck equation for $n=2$.  Dashed lines: the short-time asymptotic~(\ref{smallT2}) and the long-time asymptotic~(\ref{longTn=2}).}
\label{num4}
\end{figure}

Figure  \ref{num4} presents our numerical results for $n=2$ (where neither a complete analytic solution, nor its Laplace transform is available) alongside with  the short-time tail~(\ref{smallT2}) (without any adjustable parameter) and the long-time tail~(\ref{longTn=2}).

\section{High Dimensions}
\label{sec:high}

\begin{figure}
\centering
\includegraphics[width=7.89cm]{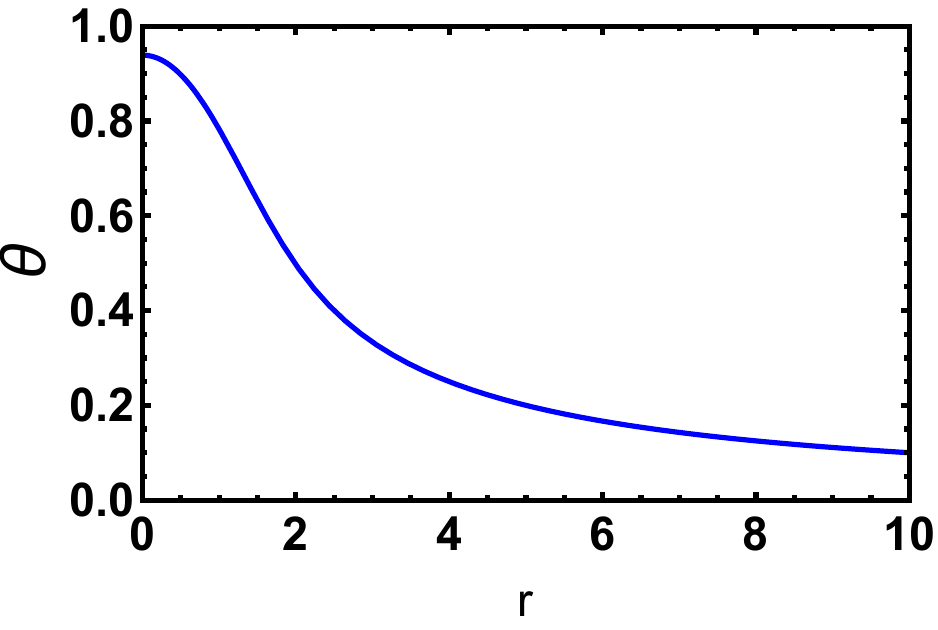}
\caption{The average rescaled blowup time versus the initial position $r$ for the cubic potential $U=-r^3/3$. The analytical solution is given by Eq.~\eqref{Tr-3}.}
\label{Fig:T3D}
\end{figure}

The average blowup time $\Theta(r)$ in $d$ dimensions satisfies the equation
\begin{equation}
\label{T-eq-r}
\frac{d^2 \Theta}{dr^2} + \frac{d-1}{r}\,\frac{d\Theta}{dr} + r^{n}\,\frac{d\Theta}{dr} = -1
\end{equation}
This equation is an ODE due to the fact that the repulsive potential is isotropic.  The derivation of \eqref{T-eq-r} is a straightforward generalization of the derivation in one dimension; for examples of such derivations and analyses of the solutions, see \textit{e.g.} Refs. \cite{SR,KRB,KR18}. The boundary conditions are
\begin{equation}
\label{BC-r}
\frac{d\Theta}{dr}\Big|_{r=0} = 0, \qquad \Theta|_{r=\infty}=0.
\end{equation}

In three dimensions, the average blowup time for the particle in a cubic potential $U=-r^3/3$, \textit{i.e.} $n=2$,  is (see also Fig.~\ref{Fig:T3D})
\begin{equation}
\label{Tr-3}
 \Theta(r) = \frac{1}{r}- 3^{-\frac{4}{3}}\,\Gamma\!\left(-\frac{1}{3}\,,
\frac{r^3}{3}\right),
\end{equation}
where $\Gamma(\ldots,\ldots)$ is the incomplete gamma function. If the particle starts at the origin, we obtain
\begin{equation}
\label{T0-3}
\langle T\rangle = \lim_{r\to 0} \Theta(r) = 3^{-\frac{1}{3}}\,\Gamma(\tfrac{2}{3})\,.
\end{equation}

Generally, the average blowup time $\Theta(r)$ decays algebraically when $r\to \infty$. More precisely
\begin{equation}
\label{av-asymp-r}
\Theta(r) = \frac{1}{(n-1)r^{n-1}}+\frac{n+1-d}{2n\,r^{2n}}+\ldots,
\end{equation}
which is most easily derived directly from Eq.~\eqref{T-eq-r}. The leading term is the deterministic blowup time. The subleading term also decays algebraically when $n+1-d\ne 0$. The above example of the particle in a cubic potential in three dimensions, $n=2$ and $d=3$, is one of the exceptional cases where the subleading term decays much faster. From the exact solution \eqref{Tr-3} we obtain in this case
\begin{equation}
\label{Tr-3-asymp}
 \Theta(r) = \frac{1}{r} - \frac{1}{r^4}\,\exp\!\left(-\frac{r^3}{3}\right)+ \frac{4}{r^7}\,\exp\!\left(-\frac{r^3}{3}\right)+\ldots
 \end{equation}
when $r\to\infty$.

We now present a few more explicit solutions. In three dimensions,  the average time admits only integral representations when $n=3$ and $n=4$. For $n=5$, one can express $T(r)$ through a hypergeometric function with four indexes:
\begin{equation}
\label{Tr-6}
\Theta(r) =  \frac{2^\frac{1}{3}\pi^\frac{3}{2}}{3^\frac{7}{6} \Gamma(\frac{1}{6})} -\tfrac{1}{6}r^2
F\big(\tfrac{1}{3},1;\tfrac{4}{3},\tfrac{3}{2}; -\tfrac{r^6}{6}\big).
\end{equation}

In two dimensions, we managed to express the average blowup time $\Theta(r)$ through the hypergeometric function only for $n=3$:
\begin{equation}
\label{Tr-4}
\Theta(r) = \tfrac{1}{8}\pi^\frac{3}{2} -\tfrac{1}{4}r^2 F\big(\tfrac{1}{2},1;\tfrac{3}{2},\tfrac{3}{2}; -\tfrac{r^2}{4}\big).
\end{equation}

The Laplace transform of the probability distribution $\mathcal{P}(T)$ of the blowup time $T$ satisfies the linear ODE
\begin{equation}
\label{Pi-eq-r}
\frac{d^2 \Pi}{dr^2} +\frac{d-1}{r}\,\frac{d\Pi}{dr} + r^{n}\,\frac{d\Pi}{dr} = s \Pi
\end{equation}
and the boundary conditions
\begin{equation}\label{BCs3d}
\frac{d \Pi}{dr}\left(r=0,s\right)=0 \quad \text{and} \quad \Pi(r\to \infty,s)=1.
\end{equation}

\begin{figure}
\centering
\includegraphics[width=0.414\textwidth,clip=]{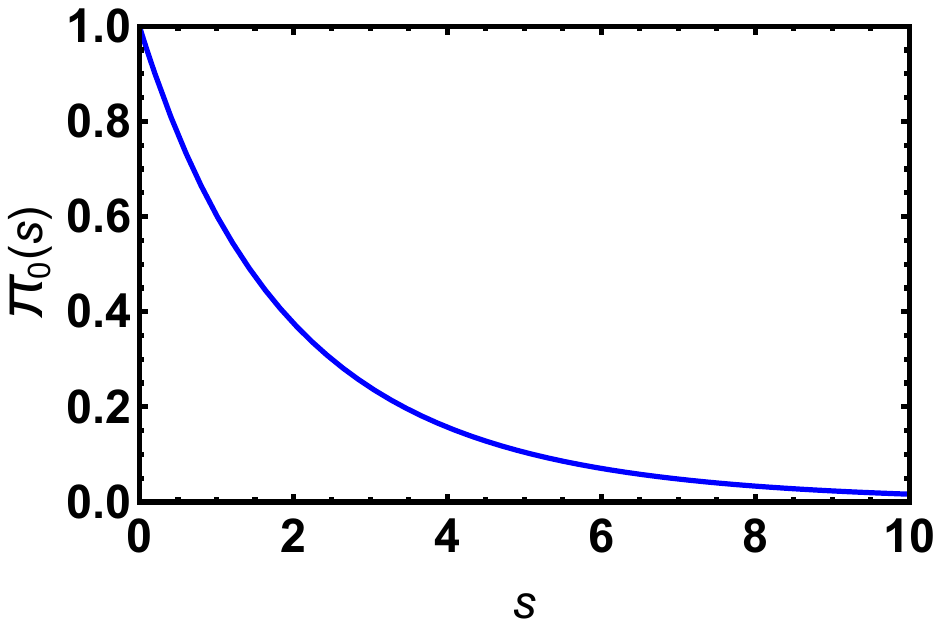}

\includegraphics[width=7.10cm]{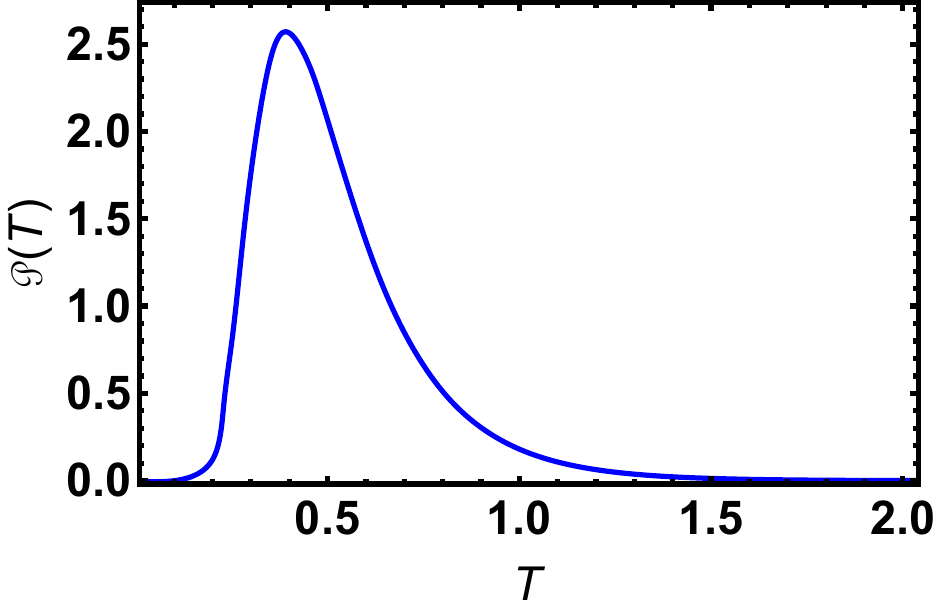}
\caption{The Laplace transform $\Pi_0(s)$ of the probability distribution $\mathcal{P}(T)$ in three dimensions for $n=3$, as described by Eq.~(\ref{Pi-0-3}) (top panel). Bottom panel: $\mathcal{P}(T)$ for $n=3$ in three dimensions obtained by numerically inverting the Laplace transform.}
\label{Fig:Pi-0-3}
\end{figure}

To solve Eq.~(\ref{Pi-eq-r}) in three dimensions we recall that the transformation $\Pi(r) = r^{-1}\Phi(r)$ reduces the three-dimensional radial Laplacian to the one-dimensional Laplacian,
\begin{equation}
\left(\frac{d^2}{dr^2} +\frac{2}{r}\,\frac{d}{dr}\right)\frac{\Phi}{r} = \frac{1}{r}\, \frac{d^2 \Phi}{dr^2}\,,
\end{equation}
and hence Eq.~(\ref{Pi-eq-r}) in three dimensions turns into
\begin{equation}
\label{Phi-3}
\frac{d^2 \Phi}{dr^2}  + r^{n}\,\frac{d\Phi}{dr} = (s+ r^{n-1}) \Phi
\end{equation}
As in one dimension, we found the solution of Eq.~\eqref{Phi-3} and the boundary conditions (\ref{BCs3d}) only for the quartic potential ($n=3$). This solution reads
\begin{equation}
\label{Pi-3-sol}
\Pi(s,r)=\frac{\text{HeunB}[s/4, 0, 3/2, 0, 1/2; r^2]}{\text{HeunB}[s/4, 0, 3/2, 0, 1/2; \infty]}.
\end{equation}
If the particle starts at the origin, the Laplace transform \eqref{Pi-3-sol} reduces to
\begin{equation}
\label{Pi-0-3}
\Pi_0(s)=\frac{1}{\text{HeunB}[s/4, 0, 3/2, 0, 1/2; \infty]};
\end{equation}
The top panel of Fig.~\ref{Fig:Pi-0-3} shows a plot of $\Pi_0(s)$ vs. $s$. The bottom panel shows the resulting probability distribution of blowup time, obtained by inverse Laplace transform of $\Pi_0(s)$.

Expression~(\ref{Pi-0-3}) has a simple pole, and hence the long-time tail of $\mathcal{P}(T)$ is purely exponential:
\begin{equation}
\mathcal{P}(T)\sim e^{-a_3 T}, \qquad a_3\simeq 5.25237.
\end{equation}
As in one dimension, $a$ is also the lowest eigenvalue of the pertinent Fokker-Planck operator, describing the evolution of the position distribution $P(r,t)$ of the particle.

The short-time tail of $\mathcal{P}(T)$ exhibits an essential singularity, again as in one dimension. The leading-order $T\to 0$ asymptotic of $\mathcal{P}(T)$, for any $n>1$, readily follows from the OFM. A crucial observation is that, since we do not condition the process on reaching a certain escape angle, the optimal path preserves the initial angle of the particle. The resulting radial Euler-Lagrange equation, $\ddot{r}(t)-nr^{2n-1}(t)=0$, coincides in all dimensions with the one-dimensional equation~(\ref{ELeq}). The boundary conditions in time, $r(0)=0$ and $r(T)=\infty$, also coincide with their one-dimensional counterparts. As a result, the optimal paths are exactly the same,  and the leading-order short-time asymptotics of $\mathcal{P}(T)$ are still described by Eqs.~(\ref{smallTgeneral})--(\ref{smallT4}). (Pre-exponential factors in these asymptotics, which we did not attempt to calculate, are expected to depend on the dimension of space.)

Inspecting the Laplace transforms of the blowup time for the particle in a quartic potential, Eq.~\eqref{Pi-sol} in one dimension and Eq.~\eqref{Pi-3-sol} in three dimensions, one can guess that, in arbitrary spatial dimension and $n=3$, the solution is
\begin{equation}
\label{Pi-d-sol}
\Pi(s,r)=\frac{\text{HeunB}[s/4, 0, d/2, 0, 1/2; r^2]}{\text{HeunB}[s/4, 0, d/2, 0, 1/2; \infty]}.
\end{equation}
Remarkably, this conjecture turns out to be correct, as can be verified by the direct substitution of \eqref{Pi-d-sol} into Eq.~(\ref{Pi-eq-r}).

The closest to the origin singularity of $\Pi(s,r)$ from Eq.~(\ref{Pi-d-sol}) is again a simple pole
\begin{equation}
\label{Pi-d-pole}
\Pi(s,r) = A_d\,\frac{\text{HeunB}[-a_d/4, 0, d/2, 0, 1/2; r^2]}{s+a_d}+\ldots
\end{equation}
so that the  long-$T$ tail of $\mathcal{P}(T)$, for any initial position of the particle $r$, is purely exponential:
\begin{equation}
\mathcal{P}(T)\simeq A_d \text{HeunB}[-a_d/4, 0, d/2, 0, 1/2; r^2]\, e^{-a_d T}\,.
\label{detailed}
\end{equation}
The characteristic decay rate $a_d$ is independent of the initial position $r$, but the rather detailed formula (\ref{detailed}) gives the $r$-dependence of the pre-exponential factor. The amplitude $A_d$ depends only on $d$. In addition to the already known decay rates $a_1$ and $a_3$, we mention the decay rates $a_2\simeq 3.14317$ and $a_4\simeq 7.65229$ in two and four dimensions.

\section{Conclusions and Discussion}
\label{conclusions}

We investigated the impact of noise on a deterministic finite-time singularity, using as an example a simple and generic model of a Brownian particle performing an overdamped motion in a power-law repulsive potential.  The average blowup time $\langle T \rangle$ can be calculated by the well-developed first-passage formalism for rather general potentials. The outcome is integral. Apart from a few exceptions, this integral does not reduce, even for power-law potentials, to an explicit formula. One exceptional case is the quartic potential, for which we derived explicit results in one, two, and three dimensions. In some dimensions, we also obtained analytical results for the average blowup time in a cubic or sextic potential. In all these solvable cases, the results are expressible through hypergeometric functions with four indexes. In one dimension, when the particle starts at the origin, we computed the average blowup time for arbitrary power-law potentials.

We also investigated the complete probability distribution $\mathcal{P}(T)$ of the blowup time. We employed three different yet complementary approaches. For the quartic potential, we expressed the Laplace transform of $\mathcal{P}(T)$  in terms of bi-confluent Heun functions in one and three dimensions. This Laplace transform has a simple pole ensuring a pure exponential large-$T$ tail of $\mathcal{P}(T)$. We also showed independently that this tail is determined by the ground state eigenmode of the pertinent Fokker-Planck equation that describes the evolution in time of the probability distribution of the particle position.

Little is known about the large $s$ behavior of the Laplace transform, which corresponds to the small-$T$ tail of $\mathcal{P}(T)$.  Fortunately, this tail is captured by the OFM. Using this method, we determined the optimal path of the process conditioned on small $T$, computed in the leading order the $T\to 0$ tail of $\mathcal{P}(T)$ and uncovered its essential singularity at $T=0$. Remarkably, this essential singularity is independent of the spatial dimension and depends only on $n$.

Comparing our one-dimensional and high-dimensional results, one observes that the effect of the spatial dimension on the blowup time is merely quantitative. New features appear for $d\geq 2$ if we are interested in the \emph{direction of escape}. Consider the two-dimensional case, and suppose that the particle starts at the point $(r,0)$ of the polar coordinates $(r,\theta)$. Just before the blowup time, the particle will be at a position $R e^{i \theta(r)}$ with $R\gg 1$. The ultimate escape angle $-\pi<\theta(r)\leq \pi$ is a random function that depends on the initial position of the particle. The distribution $P(\theta,r)$ is symmetric, $P(\theta,r)=P(-\theta,r)$, and single-peaked with a maximum at $\theta=0$. If the particle starts at $r=0$, the angle distribution is uniform. For other initial conditions, the average angle vanishes, $\langle \theta(r)\rangle=0$, while the variance $\langle [\theta(r)]^2\rangle$ is non-trivial. From dimensional analysis, $\langle [\theta(r)]^2\rangle = F(\rho)$ with $\rho=r/\ell=r(g/D)^\frac{1}{n+1}$. We know that $F(0)=\pi^2/3$ (the variance of the uniform distribution on the interval of length $2\pi$),  and we expect $F(\rho)$ to be a decreasing function of $\rho$ as the deterministic force dominates more and more over the noise.

A natural way of arriving at the blowup problem is by studying the behavior of an ensemble of particles performing independent Brownian motions and an overdamped motion caused by repulsive pairwise interactions \cite{Kirone24}. Reference \cite{Kirone24} analyzes the expansion dynamics of such a many-particle system for pairwise potentials for which a blowup is impossible. It would be interesting to consider the blowup regime as well. A two-particle system essentially reduces to the problem we studied here, and a blowup is inevitable for potentials that grow faster than quadratically. For such potentials, all $N$ particles are expected to escape to infinity simultaneously. This conjectural behavior appears obvious in $d\geq 2$ dimensions. It also seems correct in one dimension. In $d\geq 2$ dimensions, the directions of the escape to infinity are also interesting. Two particles escape in opposite directions. When $N=3$, the distribution of directions could be a non-trivial function even when all particles start at the same point.

Switching from the overdamped motion to conservative one is an intriguing challenge. In one dimension, the governing equation would be
\begin{equation}
\label{accel}
\ddot x(t) = gx^{n}(t) + \eta(t),
\end{equation}
with $\eta(t)$ being the white Gaussian noise. The blowup again occurs when $n>2$. The average blowup time $\Theta(x,v)$ for a particle with the initial position $x$ and velocity $v$ satisfies a partial differential equation
\begin{equation}
\label{Txv-eq}
D\partial_v^2 \Theta +gx^{n}\partial_v \Theta + v\partial_x \Theta = -1,
\end{equation}
rather than an ODE, Eq.~\eqref{avtimeeq}, that we had in the overdamped case. Equation \eqref{Txv-eq} with $g=0$ on a finite interval describes the average exit time for a randomly accelerated particle. Extending the corresponding exact solution \cite{Masoliver95,Masoliver96}, and several other results describing first-passage problems for the randomly accelerated particle \cite{McKean62,Marshall85,Bray98,Burk06,Majumdar10,Burk-rev,Meerson23} to the case of $g\neq 0$ requires a dedicated analysis, which we leave for the future.

\bigskip
\noindent
{\bf Acknowledgments}. We are very grateful to E. Barkai, who attracted our attention to Refs.~\cite{Ornigotti,SilerPRL,Ryabov}, and to N. R. Smith, whose advice we used in Appendix \ref{ap:nozero}. We also thank K. Mallick and S. Redner for useful discussions.  P.L.K. is grateful to IPhT (Saclay), the University of Aveiro, and the University of Granada for their hospitality.  The research of B.M. was supported by the Israel Science Foundation (Grant No. 1499/20).

\appendix
\section{Cumulants}
\label{ap:cum}

Using the Laplace transform of the blowup time, one can compute the cumulants $\langle [T(x)]^k\rangle_c$ of the blowup time by expanding the logarithm of the Laplace transform:
\begin{equation}
\label{log-cum}
\log \Pi(s,x)=\sum_{k\geq 1} \frac{(-s)^k}{k!}\,\langle [T(x)]^k\rangle_c
\end{equation}

For the Brownian particle in the quartic potential in one dimension, the Laplace transform is given by \eqref{Pi-0} if the particle starts at the origin. Using the shorthand notation
\begin{equation}
\label{B:1}
B(z)= \text{HeunB}[z, 0, 1/2, 0, 1/2; \infty],
\end{equation}
we rewrite \eqref{Pi-0} as $\Pi_0(s)=1/B(s/4)$ which, in conjunction with Eq.~\eqref{log-cum}, allows us to express the cumulants via derivatives of the function $B(z)$ taken at $z=0$. We shortly write these derivatives as $B', B''$, \textit{etc}. Taking into account that $B(0)=1$ we find that the cumulants have the form
$\langle T^n\rangle_c=4^{-n}\, \mathcal{P}_n$, where $\mathcal{P}_n$ are polynomials of the derivatives $B', B'',\ldots$ with integer coefficients:
\begin{equation*}
\label{cum-B}
\begin{split}
\mathcal{P}_1 & =  B', \\
\mathcal{P}_2 & =  (B')^2-B'', \\
\mathcal{P}_3 & =  2(B')^3-3 B' B''+ B''', \\
\mathcal{P}_4 & =  6(B')^4-12(B')^2 B''+3(B'')^2 + 4 B' B''' - B^{(4)}, \\
\mathcal{P}_5& =  24(B')^5- 60 (B')^3 B''+30 B' (B'')^2\\
                            &+ 20 (B')^2 B''' - 10 B'' B''' -5 B' B^{(4)} +  B^{(5)},
\end{split}
\end{equation*}
\textit{etc}. The polynomials $\mathcal{P}_n$ are homogeneous of degree $n$ if we assign degree $m$ to the $m^\text{th}$ derivative of the Heun function \eqref{B:1}, viz., $\deg[B^{(m)}]=m$.  Using {\em Mathematica},  one can obtain accurate numerical values for cumulants.

For the Brownian particle in the quartic potential in higher dimensions, the Laplace transform \eqref{Pi-d-sol} reduces to
\begin{equation}
\label{Pi-d-0}
\Pi_0(s)=\frac{1}{\text{HeunB}[s/4, 0, d/2, 0, 1/2; \infty]}
\end{equation}
if the particle starts at the origin.  The cumulants of the blowup time of the particle starting at the origin have again the form $\langle T^n\rangle_c=4^{-n}\, \mathcal{P}_n$ with polynomials $\mathcal{P}_n$ given by the same formulas as before. The only distinction is that
\begin{equation}
\label{B-d}
B(z)= \text{HeunB}[z,0, d/2, 0, 1/2; \infty]
\end{equation}
in the $d$-dimensional case.

\section{All eigenvalues of the Fokker-Planck operator are positive}
\label{ap:nozero}

As we have seen in Sec.~\ref{Laplace}, for even $n$ there is an interval of $x$ where the potential $V(x,n)$ is negative. Here we show that, in spite of this fact, all the eigenvalues $\lambda_k$, $k=1,2, \ldots$ are positive. It suffices to prove this statement for the ground state eigenvalue $\lambda_1$.  Let us consider an auxiliary problem by introducing hard walls at $|x|=L$. The modified potential of the Shr\"{o}dinger equation is
\begin{equation}
V_L(x,n) =
\begin{cases}
\frac{x^{2 n}}{4}+\frac{n x^{n-1}}{2}\,,                                  &|x|<L\,,\\
+\infty\,,      & |x|>L\,.
\end{cases}
\label{VL}
\end{equation}

The modified eigenvalues depend on $L$, and the original eigenvalues are recovered in the $L\to \infty$ limit. We prove that $\lambda_1(L \to \infty)>0$ by contradiction. Assume for a moment that $\lambda_1(L \to \infty)$ is negative and consider the opposite $L\to 0$ limit. Since the potential $V(x,n)$ is regular at $x=0$ and vanishes there, the leading-order asymptotic behavior of $\lambda_1(L\to 0)$ is the same as that of the ground-state eigenvalue in the infinite square well
\begin{equation}
V_0(x) =
\begin{cases}
0\,,                                  &|x|<L\,,\\
+\infty\,,      & |x|>L\,.
\end{cases}
\label{squarewell}
\end{equation}
Thus $\lambda_1(L\to 0) \sim L^{-2}$ \cite{LLQM}, and it is certainly positive. Therefore, if $\lambda_1(L \to \infty)$ is negative, there must exist an intermediate value $L=L_*$ such that $\lambda_1(L_*)=0$.  But this implies that our original Schr\"{o}dinger equation~(\ref{Schroedinger}) has a zero-energy solution which vanishes at some finite points $x = \pm L_*$ to comply with the boundary conditions
\begin{equation}
\label{BCwell}
v_1(x=\pm L)=0
\end{equation}
of the modified problem.

We now show that such a solution does not exist. Let us return to Eq.~(\ref{ODEu}) with $\lambda_1=0$:
\begin{equation}
\label{uzero}
u_1^{\prime\prime}(x) - \left[x^n u_1(x)\right]^{\prime}= 0\,.
\end{equation}
This equation in total derivatives can be easily solved. Its general solution is
\begin{equation}
\label{ugen}
u_1(x) = C_1 + C_2 \int_0^x dy\,\exp\!\left(\frac{y^{n+1}}{n+1}\right).
\end{equation}
Using the transformation~(\ref{transform1}), one can obtain the general zero-energy solution $v_1(x)$ of the Shr\"{o}dinger equation~(\ref{Schroedinger}). This solution, however, differs from $u_1(x)$ only by the factor $\exp\left[-\frac{x^{n+1}}{2(n+1)}\right]$ which is strictly positive. Therefore, the conditions~(\ref{BCwell}) are equivalent to the conditions $u_1(x=\pm L)=0$.  For even $n$, we combine \eqref{ugen} with $u_1(x=\pm L)=0$ to yield  \begin{eqnarray*}
\label{eqL}
  C_1+C_2 \int_0^L dy\,\exp\!\left(\frac{y^{n+1}}{n+1}\right) &=& 0\,,\\
 C_1 - C_2 \int_0^{L} dy\,\exp\!\left(-\frac{y^{n+1}}{n+1}\right)&=& 0\,.
\end{eqnarray*}
These equations have only a trivial solution $C_1=C_2=0$ for any $L>0$. This proves the non-existence of $L_*$ and disproves the assumption that $\lambda_1=\lambda_1(L\to \infty)$ is negative. Finally, the ground state eigenvalue $\lambda_1$ cannot be zero because the solution does not vanish at infinity.

\bibliography{vacuum}

\begin{thebibliography}{27}%
\makeatletter
\providecommand \@ifxundefined [1]{%
 \@ifx{#1\undefined}
}%
\providecommand \@ifnum [1]{%
 \ifnum #1\expandafter \@firstoftwo
 \else \expandafter \@secondoftwo
 \fi
}%
\providecommand \@ifx [1]{%
 \ifx #1\expandafter \@firstoftwo
 \else \expandafter \@secondoftwo
 \fi
}%
\providecommand \natexlab [1]{#1}%
\providecommand \enquote  [1]{``#1''}%
\providecommand \bibnamefont  [1]{#1}%
\providecommand \bibfnamefont [1]{#1}%
\providecommand \citenamefont [1]{#1}%
\providecommand \href@noop [0]{\@secondoftwo}%
\providecommand \href [0]{\begingroup \@sanitize@url \@href}%
\providecommand \@href[1]{\@@startlink{#1}\@@href}%
\providecommand \@@href[1]{\endgroup#1\@@endlink}%
\providecommand \@sanitize@url [0]{\catcode `\\12\catcode `\$12\catcode
  `\&12\catcode `\#12\catcode `\^12\catcode `\_12\catcode `\%12\relax}%
\providecommand \@@startlink[1]{}%
\providecommand \@@endlink[0]{}%
\providecommand \url  [0]{\begingroup\@sanitize@url \@url }%
\providecommand \@url [1]{\endgroup\@href {#1}{\urlprefix }}%
\providecommand \urlprefix  [0]{URL }%
\providecommand \Eprint [0]{\href }%
\providecommand \doibase [0]{http://dx.doi.org/}%
\providecommand \selectlanguage [0]{\@gobble}%
\providecommand \bibinfo  [0]{\@secondoftwo}%
\providecommand \bibfield  [0]{\@secondoftwo}%
\providecommand \translation [1]{[#1]}%
\providecommand \BibitemOpen [0]{}%
\providecommand \bibitemStop [0]{}%
\providecommand \bibitemNoStop [0]{.\EOS\space}%
\providecommand \EOS [0]{\spacefactor3000\relax}%
\providecommand \BibitemShut  [1]{\csname bibitem#1\endcsname}%
\let\auto@bib@innerbib\@empty
\bibitem [{\citenamefont {Mather}\ and\ \citenamefont
  {McGehee}(1975)}]{Mather}%
  \BibitemOpen
  \bibfield  {author} {\bibinfo {author} {\bibfnamefont {J.}~\bibnamefont
  {Mather}}\ and\ \bibinfo {author} {\bibfnamefont {R.}~\bibnamefont
  {McGehee}},\ }in\ \href {https://doi.org/10.1007/3-540-07171-7_18} {\emph
  {\bibinfo {booktitle} {Dynamical Systems, Theory and Applications}}},\
  \bibinfo {editor} {edited by\ \bibinfo {editor} {\bibfnamefont
  {J.}~\bibnamefont {Moser}}}\ (\bibinfo  {publisher} {Springer},\ \bibinfo
  {address} {Berlin},\ \bibinfo {year} {1975})\ pp.\ \bibinfo {pages}
  {573--597}\BibitemShut {NoStop}%
\bibitem [{\citenamefont {Caflisch}\ and\ \citenamefont
  {Papanicolau}(1993)}]{Caflisch}%
  \BibitemOpen
  \bibfield  {author} {\bibinfo {author} {\bibfnamefont {R.~E.}\ \bibnamefont
  {Caflisch}}\ and\ \bibinfo {author} {\bibfnamefont {G.}~\bibnamefont
  {Papanicolau}},\ }\href {https://doi.org/10.1007/978-94-011-2022-7} {\emph
  {\bibinfo {title} {Singularities in Fluids, Plasmas and Optics}}}\ (\bibinfo
  {publisher} {Kluwer},\ \bibinfo {address} {Dordrecht},\ \bibinfo {year}
  {1993})\BibitemShut {NoStop}%
\bibitem [{\citenamefont {Eggers}\ and\ \citenamefont
  {Fontelos}(2009)}]{Eggers09}%
  \BibitemOpen
  \bibfield  {author} {\bibinfo {author} {\bibfnamefont {J.}~\bibnamefont
  {Eggers}}\ and\ \bibinfo {author} {\bibfnamefont {M.~A.}\ \bibnamefont
  {Fontelos}},\ }\href {\doibase 10.1088/0951-7715/22/1/R01} {\bibfield
  {journal} {\bibinfo  {journal} {Nonlinearity}\ }\textbf {\bibinfo {volume}
  {22}},\ \bibinfo {pages} {R1} (\bibinfo {year} {2009})}\BibitemShut {NoStop}%
\bibitem [{\citenamefont {Bray}(2000)}]{Bray2000}%
  \BibitemOpen
  \bibfield  {author} {\bibinfo {author} {\bibfnamefont {A.~J.}\ \bibnamefont
  {Bray}},\ }\href {\doibase 10.1103/PhysRevE.62.103} {\bibfield  {journal}
  {\bibinfo  {journal} {Phys. Rev. E}\ }\textbf {\bibinfo {volume} {62}},\
  \bibinfo {pages} {103} (\bibinfo {year} {2000})}\BibitemShut {NoStop}%
\bibitem [{\citenamefont {Farago}(2000)}]{Farago2000}%
  \BibitemOpen
  \bibfield  {author} {\bibinfo {author} {\bibfnamefont {J.}~\bibnamefont
  {Farago}},\ }\href {\doibase 10.1088/0951-7715/22/1/R01} {\bibfield
  {journal} {\bibinfo  {journal} {Europhys. Lett.}\ }\textbf {\bibinfo {volume}
  {52}},\ \bibinfo {pages} {379} (\bibinfo {year} {2000})}\BibitemShut
  {NoStop}%
\bibitem [{\citenamefont {Fogedby}\ and\ \citenamefont
  {Poutkaradze}(2002)}]{Fogedby2002}%
  \BibitemOpen
  \bibfield  {author} {\bibinfo {author} {\bibfnamefont {H.~C.}\ \bibnamefont
  {Fogedby}}\ and\ \bibinfo {author} {\bibfnamefont {V.}~\bibnamefont
  {Poutkaradze}},\ }\href {\doibase 10.1088/0305-4470/39/41/S07} {\bibfield
  {journal} {\bibinfo  {journal} {Phys. Rev. E}\ }\textbf {\bibinfo {volume}
  {66}},\ \bibinfo {pages} {021103} (\bibinfo {year} {2002})}\BibitemShut
  {NoStop}%
\bibitem [{\citenamefont {Ornigotti}\ \emph {et~al.}(2018)\citenamefont
  {Ornigotti}, \citenamefont {Ryabov}, \citenamefont {Holubec},\ and\
  \citenamefont {Filip}}]{Ornigotti}%
  \BibitemOpen
  \bibfield  {author} {\bibinfo {author} {\bibfnamefont {L.}~\bibnamefont
  {Ornigotti}}, \bibinfo {author} {\bibfnamefont {A.}~\bibnamefont {Ryabov}},
  \bibinfo {author} {\bibfnamefont {V.}~\bibnamefont {Holubec}}, \ and\
  \bibinfo {author} {\bibfnamefont {R.}~\bibnamefont {Filip}},\ }\href
  {\doibase 10.1103/PhysRevE.97.032127} {\bibfield  {journal} {\bibinfo
  {journal} {Phys. Rev. E}\ }\textbf {\bibinfo {volume} {97}},\ \bibinfo
  {pages} {032127} (\bibinfo {year} {2018})}\BibitemShut {NoStop}%
\bibitem [{\citenamefont {\v{S}iler}\ \emph {et~al.}(2018)\citenamefont
  {\v{S}iler}, \citenamefont {Ornigotti}, \citenamefont {Brzobohat\'{y}},
  \citenamefont {J\'{a}kl}, \citenamefont {Ryabov}, \citenamefont {Holubec},
  \citenamefont {Zem\'{a}nek},\ and\ \citenamefont {Filip}}]{SilerPRL}%
  \BibitemOpen
  \bibfield  {author} {\bibinfo {author} {\bibfnamefont {M.}~\bibnamefont
  {\v{S}iler}}, \bibinfo {author} {\bibfnamefont {L.}~\bibnamefont
  {Ornigotti}}, \bibinfo {author} {\bibfnamefont {O.}~\bibnamefont
  {Brzobohat\'{y}}}, \bibinfo {author} {\bibfnamefont {P.}~\bibnamefont
  {J\'{a}kl}}, \bibinfo {author} {\bibfnamefont {A.}~\bibnamefont {Ryabov}},
  \bibinfo {author} {\bibfnamefont {V.}~\bibnamefont {Holubec}}, \bibinfo
  {author} {\bibfnamefont {P.}~\bibnamefont {Zem\'{a}nek}}, \ and\ \bibinfo
  {author} {\bibfnamefont {R.}~\bibnamefont {Filip}},\ }\href {\doibase
  10.1103/PhysRevLett.121.230601} {\bibfield  {journal} {\bibinfo  {journal}
  {Phys. Rev. Lett.}\ }\textbf {\bibinfo {volume} {121}},\ \bibinfo {pages}
  {230601} (\bibinfo {year} {2018})}\BibitemShut {NoStop}%
\bibitem [{\citenamefont {Ryabov}\ \emph {et~al.}(2019)\citenamefont {Ryabov},
  \citenamefont {Holubec},\ and\ \citenamefont {Berestneva}}]{Ryabov}%
  \BibitemOpen
  \bibfield  {author} {\bibinfo {author} {\bibfnamefont {A.}~\bibnamefont
  {Ryabov}}, \bibinfo {author} {\bibfnamefont {V.}~\bibnamefont {Holubec}}, \
  and\ \bibinfo {author} {\bibfnamefont {E.}~\bibnamefont {Berestneva}},\
  }\href {\doibase 10.1103/PhysRevLett.121.230601} {\bibfield  {journal}
  {\bibinfo  {journal} {J. Stat. Mech.}\ }\textbf {\bibinfo {volume} {121}},\
  \bibinfo {pages} {084014} (\bibinfo {year} {2019})}\BibitemShut {NoStop}%
\bibitem [{\citenamefont {Hirsch}\ \emph {et~al.}(1982)\citenamefont {Hirsch},
  \citenamefont {Huberman},\ and\ \citenamefont {Scalapino}}]{Hirsch}%
  \BibitemOpen
  \bibfield  {author} {\bibinfo {author} {\bibfnamefont {J.~E.}\ \bibnamefont
  {Hirsch}}, \bibinfo {author} {\bibfnamefont {B.~A.}\ \bibnamefont
  {Huberman}}, \ and\ \bibinfo {author} {\bibfnamefont {D.~J.}\ \bibnamefont
  {Scalapino}},\ }\href {\doibase 10.1103/PhysRevA.25.519} {\bibfield
  {journal} {\bibinfo  {journal} {Phys. Rev. A}\ }\textbf {\bibinfo {volume}
  {25}},\ \bibinfo {pages} {519} (\bibinfo {year} {1982})}\BibitemShut
  {NoStop}%
\bibitem [{\citenamefont {Sigeti}\ and\ \citenamefont
  {Horsthemke}(1989)}]{Horsthemke}%
  \BibitemOpen
  \bibfield  {author} {\bibinfo {author} {\bibfnamefont {D.}~\bibnamefont
  {Sigeti}}\ and\ \bibinfo {author} {\bibfnamefont {W.}~\bibnamefont
  {Horsthemke}},\ }\href {https://doi.org/10.1007/BF01044713} {\bibfield
  {journal} {\bibinfo  {journal} {J. Stat. Phys.}\ }\textbf {\bibinfo {volume}
  {54}},\ \bibinfo {pages} {1217} (\bibinfo {year} {1989})}\BibitemShut
  {NoStop}%
\bibitem [{\citenamefont {H\"{a}nggi}\ \emph {et~al.}(1990)\citenamefont
  {H\"{a}nggi}, \citenamefont {Talkner},\ and\ \citenamefont
  {Borkovec}}]{Hanggi}%
  \BibitemOpen
  \bibfield  {author} {\bibinfo {author} {\bibfnamefont {P.}~\bibnamefont
  {H\"{a}nggi}}, \bibinfo {author} {\bibfnamefont {P.}~\bibnamefont {Talkner}},
  \ and\ \bibinfo {author} {\bibfnamefont {M.}~\bibnamefont {Borkovec}},\
  }\href {https://doi.org/10.1103/RevModPhys.62.251} {\bibfield  {journal}
  {\bibinfo  {journal} {Rev. Mod. Phys.}\ }\textbf {\bibinfo {volume} {62}},\
  \bibinfo {pages} {251} (\bibinfo {year} {1990})}\BibitemShut {NoStop}%
\bibitem [{\citenamefont {Redner}(2001)}]{SR}%
  \BibitemOpen
  \bibfield  {author} {\bibinfo {author} {\bibfnamefont {S.}~\bibnamefont
  {Redner}},\ }\href {https://doi.org/10.1017/CBO9780511606014} {\emph
  {\bibinfo {title} {A Guide to First-Passage Processes}}}\ (\bibinfo
  {publisher} {Cambridge University Press},\ \bibinfo {address} {Cambridge,
  UK},\ \bibinfo {year} {2001})\BibitemShut {NoStop}%
\bibitem [{\citenamefont {Krapivsky}\ \emph {et~al.}(2005)\citenamefont
  {Krapivsky}, \citenamefont {Redner},\ and\ \citenamefont {Ben-Naim}}]{KRB}%
  \BibitemOpen
  \bibfield  {author} {\bibinfo {author} {\bibfnamefont {P.~L.}\ \bibnamefont
  {Krapivsky}}, \bibinfo {author} {\bibfnamefont {S.}~\bibnamefont {Redner}}, \
  and\ \bibinfo {author} {\bibfnamefont {E.}~\bibnamefont {Ben-Naim}},\ }\href
  {https://doi.org/10.1017/CBO9780511780516} {\emph {\bibinfo {title} {A
  Kinetic View of Statistical Physics}}}\ (\bibinfo  {publisher} {Cambridge
  University Press},\ \bibinfo {address} {Cambridge, UK},\ \bibinfo {year}
  {2005})\BibitemShut {NoStop}%
\bibitem [{\citenamefont {Krapivsky}\ and\ \citenamefont
  {Redner}(2018)}]{KR18}%
  \BibitemOpen
  \bibfield  {author} {\bibinfo {author} {\bibfnamefont {P.~L.}\ \bibnamefont
  {Krapivsky}}\ and\ \bibinfo {author} {\bibfnamefont {S.}~\bibnamefont
  {Redner}},\ }\href {\doibase 10.1088/1742-5468/aaddb3} {\bibfield  {journal}
  {\bibinfo  {journal} {J. Stat. Mech.}\ }\textbf {\bibinfo {volume} {2018}},\
  \bibinfo {pages} {093208} (\bibinfo {year} {2018})}\BibitemShut {NoStop}%
\bibitem [{\citenamefont {Olver}\ \emph {et~al.}(2010)\citenamefont {Olver},
  \citenamefont {Lozier}, \citenamefont {Boisvert},\ and\ \citenamefont
  {Clark}}]{NIST}%
  \BibitemOpen
  \bibfield  {author} {\bibinfo {author} {\bibfnamefont {F.~W.~J.}\
  \bibnamefont {Olver}}, \bibinfo {author} {\bibfnamefont {D.~W.}\ \bibnamefont
  {Lozier}}, \bibinfo {author} {\bibfnamefont {R.~F.}\ \bibnamefont
  {Boisvert}}, \ and\ \bibinfo {author} {\bibfnamefont {C.~W.}\ \bibnamefont
  {Clark}},\ }\href {https://dlmf.nist.gov} {\emph {\bibinfo {title} {NIST
  Handbook of Mathematical Functions}}}\ (\bibinfo  {publisher} {Cambridge
  University Press},\ \bibinfo {address} {Cambridge, UK},\ \bibinfo {year}
  {2010})\BibitemShut {NoStop}%
\bibitem [{\citenamefont {Krapivsky}\ and\ \citenamefont
  {Mallick}(2024)}]{Kirone24}%
  \BibitemOpen
  \bibfield  {author} {\bibinfo {author} {\bibfnamefont {P.~L.}\ \bibnamefont
  {Krapivsky}}\ and\ \bibinfo {author} {\bibfnamefont {K.}~\bibnamefont
  {Mallick}},\ }\href {https://arxiv.org/abs/2412.14875} {\bibfield  {journal}
  {\bibinfo  {journal} {arXiv:2412.14875}\ } (\bibinfo {year}
  {2024})}\BibitemShut {NoStop}%
\bibitem [{\citenamefont {Masoliver}\ and\ \citenamefont
  {Porr\`a}(1995)}]{Masoliver95}%
  \BibitemOpen
  \bibfield  {author} {\bibinfo {author} {\bibfnamefont {J.}~\bibnamefont
  {Masoliver}}\ and\ \bibinfo {author} {\bibfnamefont {J.~M.}\ \bibnamefont
  {Porr\`a}},\ }\href {\doibase 10.1103/PhysRevLett.75.189} {\bibfield
  {journal} {\bibinfo  {journal} {Phys. Rev. Lett.}\ }\textbf {\bibinfo
  {volume} {75}},\ \bibinfo {pages} {189} (\bibinfo {year} {1995})}\BibitemShut
  {NoStop}%
\bibitem [{\citenamefont {Masoliver}\ and\ \citenamefont
  {Porr\`a}(1996)}]{Masoliver96}%
  \BibitemOpen
  \bibfield  {author} {\bibinfo {author} {\bibfnamefont {J.}~\bibnamefont
  {Masoliver}}\ and\ \bibinfo {author} {\bibfnamefont {J.~M.}\ \bibnamefont
  {Porr\`a}},\ }\href {\doibase 10.1103/PhysRevE.53.2243} {\bibfield  {journal}
  {\bibinfo  {journal} {Phys. Rev. E}\ }\textbf {\bibinfo {volume} {53}},\
  \bibinfo {pages} {2243} (\bibinfo {year} {1996})}\BibitemShut {NoStop}%
\bibitem [{\citenamefont {McKean}(1962)}]{McKean62}%
  \BibitemOpen
  \bibfield  {author} {\bibinfo {author} {\bibfnamefont {H.~P.}\ \bibnamefont
  {McKean}},\ }\href {\doibase 10.1215/kjm/1250524936} {\bibfield  {journal}
  {\bibinfo  {journal} {J. Math. Kyoto Univ.}\ }\textbf {\bibinfo {volume}
  {2}},\ \bibinfo {pages} {227} (\bibinfo {year} {1962})}\BibitemShut {NoStop}%
\bibitem [{\citenamefont {Marshall}\ and\ \citenamefont
  {Watson}(1985)}]{Marshall85}%
  \BibitemOpen
  \bibfield  {author} {\bibinfo {author} {\bibfnamefont {T.~W.}\ \bibnamefont
  {Marshall}}\ and\ \bibinfo {author} {\bibfnamefont {E.~J.}\ \bibnamefont
  {Watson}},\ }\href {\doibase 10.1088/0305-4470/18/18/016} {\bibfield
  {journal} {\bibinfo  {journal} {J. Phys. A}\ }\textbf {\bibinfo {volume}
  {18}},\ \bibinfo {pages} {3531} (\bibinfo {year} {1985})}\BibitemShut
  {NoStop}%
\bibitem [{\citenamefont {Cornell}\ \emph {et~al.}(1998)\citenamefont
  {Cornell}, \citenamefont {Swift},\ and\ \citenamefont {Bray}}]{Bray98}%
  \BibitemOpen
  \bibfield  {author} {\bibinfo {author} {\bibfnamefont {S.~J.}\ \bibnamefont
  {Cornell}}, \bibinfo {author} {\bibfnamefont {M.~R.}\ \bibnamefont {Swift}},
  \ and\ \bibinfo {author} {\bibfnamefont {A.~J.}\ \bibnamefont {Bray}},\
  }\href {\doibase 10.1103/PhysRevLett.81.1142} {\bibfield  {journal} {\bibinfo
   {journal} {Phys. Rev. Lett.}\ }\textbf {\bibinfo {volume} {81}},\ \bibinfo
  {pages} {1142} (\bibinfo {year} {1998})}\BibitemShut {NoStop}%
\bibitem [{\citenamefont {Burkhardt}\ and\ \citenamefont
  {Kotsev}(2006)}]{Burk06}%
  \BibitemOpen
  \bibfield  {author} {\bibinfo {author} {\bibfnamefont {T.~W.}\ \bibnamefont
  {Burkhardt}}\ and\ \bibinfo {author} {\bibfnamefont {S.~N.}\ \bibnamefont
  {Kotsev}},\ }\href {\doibase 10.1103/PhysRevE.73.046121} {\bibfield
  {journal} {\bibinfo  {journal} {Phys. Rev. E}\ }\textbf {\bibinfo {volume}
  {73}},\ \bibinfo {pages} {046121} (\bibinfo {year} {2006})}\BibitemShut
  {NoStop}%
\bibitem [{\citenamefont {Majumdar}\ \emph {et~al.}(2010)\citenamefont
  {Majumdar}, \citenamefont {Rosso},\ and\ \citenamefont {Zoia}}]{Majumdar10}%
  \BibitemOpen
  \bibfield  {author} {\bibinfo {author} {\bibfnamefont {S.~N.}\ \bibnamefont
  {Majumdar}}, \bibinfo {author} {\bibfnamefont {A.}~\bibnamefont {Rosso}}, \
  and\ \bibinfo {author} {\bibfnamefont {A.}~\bibnamefont {Zoia}},\ }\href
  {\doibase 10.1088/1751-8113/43/11/115001} {\bibfield  {journal} {\bibinfo
  {journal} {J. Phys. A}\ }\textbf {\bibinfo {volume} {43}},\ \bibinfo {pages}
  {115001} (\bibinfo {year} {2010})}\BibitemShut {NoStop}%
\bibitem [{\citenamefont {Burkhardt}(2014)}]{Burk-rev}%
  \BibitemOpen
  \bibfield  {author} {\bibinfo {author} {\bibfnamefont {T.~W.}\ \bibnamefont
  {Burkhardt}},\ }in\ \href {https://doi.org/10.1142/9104} {\emph {\bibinfo
  {booktitle} {First-Passage Phenomena and Their Applications}}},\ \bibinfo
  {editor} {edited by\ \bibinfo {editor} {\bibfnamefont {R.}~\bibnamefont
  {Metzler}}, \bibinfo {editor} {\bibfnamefont {G.}~\bibnamefont {Oshanin}}, \
  and\ \bibinfo {editor} {\bibfnamefont {S.}~\bibnamefont {Redner}}}\ (\bibinfo
   {publisher} {World Scientific},\ \bibinfo {address} {Singapore},\ \bibinfo
  {year} {2014})\ pp.\ \bibinfo {pages} {21--44}\BibitemShut {NoStop}%
\bibitem [{\citenamefont {Meerson}(2023)}]{Meerson23}%
  \BibitemOpen
  \bibfield  {author} {\bibinfo {author} {\bibfnamefont {B.}~\bibnamefont
  {Meerson}},\ }\href {\doibase 10.1103/PhysRevE.107.064122} {\bibfield
  {journal} {\bibinfo  {journal} {Phys. Rev. E}\ }\textbf {\bibinfo {volume}
  {107}},\ \bibinfo {pages} {064122} (\bibinfo {year} {2023})}\BibitemShut
  {NoStop}%
\bibitem [{\citenamefont {Landau}\ and\ \citenamefont {Lifshitz}(1991)}]{LLQM}%
  \BibitemOpen
  \bibfield  {author} {\bibinfo {author} {\bibfnamefont {L.~D.}\ \bibnamefont
  {Landau}}\ and\ \bibinfo {author} {\bibfnamefont {E.~M.}\ \bibnamefont
  {Lifshitz}},\ }\href@noop {} {\emph {\bibinfo {title} {Quantum Mechanics:
  Non-Relativistic Theory}}}\ (\bibinfo  {publisher} {Pergamon Press},\
  \bibinfo {address} {Oxford, UK},\ \bibinfo {year} {1991})\BibitemShut
  {NoStop}%
\end{thebibliography}%

\end{document}